\documentclass[11pt]{article} 

\usepackage[letterpaper,margin=1in]{geometry}

\usepackage[T1]{fontenc}
\usepackage{microtype}
\usepackage{wrapfig}
\usepackage{algorithm}

\usepackage{titlesec}
\titleformat*{\paragraph}{\bfseries}

\usepackage{amsthm,mathtools}
\usepackage{xcolor}
\usepackage{empheq}
\usepackage{dsfont}
\usepackage{mathrsfs}
\usepackage{graphicx}
\usepackage{enumitem}
\usepackage{verbatim}
\usepackage{aliascnt} 
\usepackage{xspace}
\usepackage{booktabs}
\usepackage{bbm}

\usepackage{caption}
\usepackage{tikz}
\usetikzlibrary{patterns}
\usetikzlibrary{arrows,shapes,automata,backgrounds,petri,positioning}
\usetikzlibrary{shadows}
\usetikzlibrary{calc}
\usetikzlibrary{spy}
\usetikzlibrary{angles, quotes}
\usetikzlibrary{matrix}
\usepackage{pgf,pgfplots}
\usepackage{pgfmath,pgffor}
\pgfplotsset{compat=1.17}

\usepackage{framed}

\definecolor[named]{ACMBlue}{cmyk}{1,0.1,0,0.1}
\definecolor[named]{ACMYellow}{cmyk}{0,0.16,1,0}
\definecolor[named]{ACMOrange}{cmyk}{0,0.42,1,0.01}
\definecolor[named]{ACMRed}{cmyk}{0,0.90,0.86,0}
\definecolor[named]{ACMLightBlue}{cmyk}{0.49,0.01,0,0}
\definecolor[named]{ACMGreen}{cmyk}{0.20,0,1,0.19}
\definecolor[named]{ACMPurple}{cmyk}{0.55,1,0,0.15}
\definecolor[named]{ACMDarkBlue}{cmyk}{1,0.58,0,0.21}

\usepackage[colorlinks,citecolor=blue,linkcolor=magenta,bookmarks=true]{hyperref}
\usepackage[nameinlink]{cleveref}
\crefname{ineq}{Inequality}{Inequality}
\creflabelformat{ineq}{#2{\upshape(#1)}#3}
\crefname{sub}{Subsection}{Subsection}
\creflabelformat{Subsection}{#2{\upshape(#1)}#3}
\crefname{sdp}{SDP}{SDP}
\creflabelformat{sdp}{#2{\upshape(#1)}#3}
\crefname{lp}{LP}{LP}
\creflabelformat{lp}{#2{\upshape(#1)}#3}
\usepackage{aliascnt}
\usepackage{cleveref}
\crefname{ineq}{Inequality}{Inequality}
\creflabelformat{ineq}{#2{\upshape(#1)}#3}
\crefname{sub}{Subsection}{Subsection}
\creflabelformat{Subsection}{#2{\upshape(#1)}#3}
\crefname{sdp}{SDP}{SDP}
\creflabelformat{sdp}{#2{\upshape(#1)}#3}
\crefname{lp}{LP}{LP}
\creflabelformat{lp}{#2{\upshape(#1)}#3}



\newtheorem{theorem}{Theorem}[section]

\newtheorem{lemma}[theorem]{Lemma}
\newtheorem{informal theorem}[theorem]{Theorem (informal statement)}

\newtheorem{corollary}[theorem]{Corollary}
\newtheorem{claim}[theorem]{Claim}
\newtheorem{fact}[theorem]{Fact}

\newtheorem{definition}[theorem]{Definition}

\DeclareMathOperator*{\E}{\mathbb{E}}

\newcommand{\F}{\mathbb{F}}
\newcommand{\supp}{\mathrm{supp}}

\def\multichoose#1#2{\ensuremath{\left(\kern-.3em\left(\genfrac{}{}{0pt}{}{#1}{#2}\right)\kern-.3em\right)}}

\newcommand{\dtv}{d_{\mathrm TV}}
\newcommand{\poly}{\mathrm{poly}}

\newcommand{\rank}{\mathrm{rank}}

\newcommand{\qcb}{\mathrm{QuadCB}}
\newcommand{\cube}[1]{\{0,1\}^{#1}}
\newcommand{\bcube}[1]{{\mathbb{F}^{#1}_2}}

\newcommand{\ext}{\mathrm{Ext}}

\newcommand{\hinf}{H_{\infty}}

\crefname{algocf}{Algorithm}{Algorithms} \Crefname{algocf}{Algorithm}{Algorithms}

\newcommand{\CB}{\mathrm{CB}}

\newcommand{\vecspan}{\mathrm{span}}

\newcommand{\mindeg}{\mathrm{mindeg}}

\newcommand{\acb}{\mathrm{AffCB}}

\newcommand{\qext}{\mathrm{QExt}}

\newcommand{\lext}{\mathrm{LExt}}

\newcommand{\nc}{\mathrm{NC}^0}
\newcommand{\ac}{\mathrm{AC}^0}
\newcommand{\nobfext}{\mathrm{NOBFExt}}

\newcommand{\rcombext}{\mathrm{RCombExt}}

\newcommand{\qdisp}{\mathrm{QDisp}}
\newcommand{\cglext}{\mathrm{CGLExt}}
\newcommand{\affinext}{\mathrm{AffineExt}}
\newcommand{\ctpartition}{\mathrm{CTPartition}}
\usepackage{newtxtext}
\usepackage{newtxmath} 

\usepackage{algorithmic}
\usepackage{multirow}
\usepackage{diagbox,xcolor}
\title{Quadratic Extractors}
\author{Gautam Chandrasekaran\thanks{This work was done in part while the authors were visiting the Simons Institute for the Theory of Computing, Berkeley.} \thanks{University of Texas at Austin. \texttt{gautamc@cs.utexas.edu}. Supported in part by the NSF AI Institute for Foundations of Machine Learning (IFML) and a Simons Investigator Award (\#409864, David Zuckerman).} 
\and Jesse Goodman\thanks{Simons Institute for the Theory of Computing. \texttt{jpmgoodman@berkeley.edu}. Most of this work was done while the author was at The University of Texas at Austin, supported by a Simons Investigator Award (\#409864, David Zuckerman).} 
\and David Zuckerman\footnotemark[1] \thanks{University of Texas at Austin. \texttt{diz@cs.utexas.edu}. Supported in part by NSF Grant CCF-2312573 and a Simons Investigator Award (\#409864).} 
}
\date{}

\begin{document}
\pagenumbering{gobble}
\maketitle
\begin{abstract}

A quadratic source is a random variable $X\sim\mathbb{F}_2^n$ generated by applying $n$ quadratics to a hidden uniform string. Previously, the best-known explicit extractor for this model required min-entropy roughly $k=n-\sqrt{\log n}$. In this paper, we give an explicit construction for min-entropy $k=\mathrm{polylog}(n)$. Even for the special case of $2$-local sources, where each bit of $X$ depends on two hidden uniform bits, achieving an entropy requirement below $k=\sqrt{n}$ was an important open problem. This is because such sources provably cannot be reduced (in a black-box manner) to our most powerful general-purpose seedless extractor, known as a sumset extractor. Our new extractor bypasses this barrier, providing the first explicit construction to go beyond sumset extraction.

\end{abstract}
\newpage
\pagenumbering{arabic}
\section{Introduction}
Uniform random bits are a fundamental building block of randomized algorithms, cryptography, and distributed computing. However, the sources of randomness we encounter in nature (usually arising from physical phenomena) are imperfect, and almost never truly uniform. The goal of the field of \textit{randomness extraction} is to design efficient deterministic algorithms that convert these imperfect (also called \textit{weak}) sources of randomness into truly uniform random bits.
\begin{definition}[Randomness Extractor]
    A randomness extractor $\ext:\{0,1\}^{n}\to \{0,1\}^m$ with error \(\epsilon\) for a class of distributions $\mathcal{X}$ over $\{0,1\}^n$ is a deterministic function such that for all $X\in\mathcal{X}$,
    \[
    \ext(X)\approx_\epsilon U_m,
    \]
    where \(\approx_\epsilon\) denotes statistical distance and $U_m$ is the uniform distribution on $m$ bits. We say that the extractor is \textit{explicit} if it is a deterministic polynomial time algorithm.
\end{definition}

We quantify the randomness present in these \textit{weak} sources using the notion of \textit{min-entropy}.
\begin{definition}[Min-Entropy]
    The min-entropy of a random variable $X$ is defined as
    \[
    \hinf(X)=\min_{x\in \supp(X)}\log_2\left(\frac{1}{\Pr[X=x]}\right).
    \]
\end{definition}

The dream randomness extractor would be a single deterministic polynomial time algorithm that works for the class of all weak random sources that have an entropy guarantee. However, this is known to be impossible. For any candidate extractor, an adversary can design a weak random source (even with min-entropy $n-1$) on which the extractor outputs a constant string. 

The community has broadly studied three approaches of extractors that get around this impossibility result. The first approach, called \textit{seeded extraction}, assumes that the extractor is also given a short truly uniform string in addition to the weak random source \cite{NZ96,Tre01,LRVW03,GUV09,DKSS13}. The second approach, called \textit{independent source extraction}, assumes that the extractor is given multiple independent weak sources as input \cite{Vaz87a,CG88,BIW06,Bou05,Li15a,CZ19,li2023two}. The final approach, and the one we study, assumes algebraic or complexity theoretic structure in the source.

\paragraph*{Algebraic Sources}
The most basic type of algebraic source is an \textit{affine source}. These are distributions obtained by passing uniform bits through an affine map (or equivalently, uniform over the pre-images of linear maps). That is, they are sources of the form $X=\ell(U_s)+b$ where $\ell:\F_2^{s}\to\F_2^{n}$ is a linear map, $U_s$ is the uniform distribution over $\F_2^{s}$ and $b\in \F_2^{n}$ is a fixed vector.  Extractors and dispersers\footnote{A disperser is a function that is non-constant on the source.} for affine sources have a rich history, starting from \cite{BHRVW01,BKSSW10,Bou07,rao2009extractors,Yeh11,BK12,Li11a,shaltiel_affine,li2016improved,chattopadhyay2021affine,chattopadhyay2022extractors} culminating in constructions achieving the optimal entropy requirement $k=O(\log n)$ by Li \cite{li2023two}. Affine extractors have also been constructed over larger fields by \cite{GR08,DG10,BDL16,BGLZ15,GVJZ23}. Affine extractors and dispersers have had various applications both within extractors \cite{viola_circuit_sources,chattopadhyay2022improved} and to circuit lower bounds \cite{DK11,CS16b,FGHK16,LY22}.

There are two natural ways to generalize affine sources to polynomials. The first is the class of \textit{variety sources}: uniform distributions over the solutions of polynomial equations. This type of source, introduced by Dvir \cite{Dvi12}, has good extractors over large fields \cite{Dvi12}, but the best constructions over $\F_2$ require entropy $0.99n$ \cite{cohen-tal,Rem16,LZ19}. Better constructions for these sources would imply new circuit lower bounds \cite{GKW21}. The second generalization is the class of \textit{polynomial sources}, which will be the main object of our study.

Polynomial sources are sources obtained by passing uniform bits through polynomial maps. That is, the sources are of the form $X=p(U_s)$ where $p$ is a low degree polynomial map. This type of source was first introduced by Dvir, Gabizon and Wigderson \cite{dvir2009extractors}, where they constructed good extractors over fields with size polynomial in $n$. Subsequently, Ben-Sasson and Gabizon \cite{gabizon_bensasson} improved their result, in particular constructing extractors for sources with linear entropy over constant size fields (at least 4). However, the techniques of both these papers, which use character sum estimates, all break down over $\F_2$. The problem of constructing \textit{quadratic source} (polynomial sources with degree 2 polynomial maps) extractors over $\F_2$ remained open for a long time, even for entropy $n-1$! Recently, Chattopadhyay, Goodman and Gurumukhani \cite{polysourceextcgg} constructed such extractors for entropy roughly $n-\sqrt{\log n}$. In this paper, we give an exponential improvement by constructing quadratic extractors over $\F_2$ with entropy requirement $\poly\log n$. 
\begin{theorem}[Extractors for quadratic sources over $\F_2$, see \Cref{thm:main_theorem_quadratic}]
\label{thm:main_intro}
There exists a universal constant $C>0$ such that for any $k\geq \log^C n$, there exists an explicit function $\qext:\{0,1\}^{n}\to \{0,1\}^m$ such that for any quadratic source $X$ over $\F_2$ with entropy at least $k$, it holds that 
\[
\qext(X)\approx_{n^{-\Omega(1)}}U_m
\]
where $m=k^{\Omega(1)}$.
\end{theorem}
Our construction is the first to go beyond sumset extraction, due to known barrier results on using sumset extractors for quadratic sources \cite{polysourceextcgg,alrabiah2024low}. Our key technical ideas include a win-win argument, a structure theorem for quadratic sources and a new object in algebraic pseudorandomness (that we call a rank combiner). We refer to \Cref{sec:technical_overview} for more details. We now discuss some implications of our result for extracting from complexity-theoretic sources.

\paragraph*{Complexity-Theoretic Sources} The other commonly studied class of structured sources involves sources motivated by complexity theory. Trevisan and Vadhan initiated this line of work in their seminal paper \cite{trevisan_vadhan} where they studied sources samplable by polynomial sized circuits. Their extractor is conditional, relying on strong complexity theoretic assumptions. The pursuit of extractors for \textit{samplable sources} under various complexity theoretic assumptions has seen remarkable progress recently \cite{BDGM23,Sha25a,BSS25,Sha25b,OS25}. Unconditional extractors are known for sources generated in small space \cite{extractors_for_small_space_kamp_zuckerman, CL16,chattopadhyay2022improved,chattopadhyay2022extractors} .

Our focus will also be on unconditional extractors. De and Watson \cite{de_watson}, and Viola \cite{viola_circuit_sources}, introduce the notion of locally samplable sources: sources obtained by passing uniform bits through local maps. A $d$-local source has the form $X=f(U_s)$, where each output bit $f_i$ of $f$ only depends on $d$ input bits. A $1$-local source (also called \textit{projections} in \cite{paturi_saks_zane}) is a special case of affine sources and hence we know explicit extractors \cite{li2023two} even for entropy $k=O(\log n)$. The situation is much worse for $2$-local sources, where the best known extractor requires entropy roughly $\sqrt{n}$ \cite{li2016improved} (by a reduction to affine sources). There were barriers (similar to the quadratic source case) on going below $\sqrt{n}$ using sumset extractors \cite{alrabiah_et_al:LIPIcs.ICALP.2022.10,alrabiah2024low}. As an immediate corollary of our main theorem \Cref{thm:main_intro}, we have the following result.
\begin{corollary}[Extractors for \(2\)-local sources]
There exists a universal constant $C>0$ such that for any $k\geq \log^C n$, there exists an explicit function $\qext:\{0,1\}^{n}\to \{0,1\}^m$ such that for any $2$-local source $X$ over $\F_2$ with entropy at least $k$, it holds that 
\[
\qext(X)\approx_{n^{-\Omega(1)}}U_m
\]
where $m=k^{\Omega(1)}$.
\end{corollary}
Our final result concerns distributions samplable by polynomial size circuits of constant depth ($\ac$). Viola \cite{viola_circuit_sources} had shown reductions from this class to local sources. He gave extractors with entropy requirement roughly $n^{2/3}$ and Li \cite{li2016improved} improved this to $\sqrt{n}$. In this work, we give an improved reduction to $2$-local sources and obtain extractors that only need entropy roughly $n^{1/3}$.
\begin{corollary}[Extractors for \(\ac\) sources, see \Cref{thm:ext_ac0}]
For any constant $\alpha>0$ and any $k\geq n^{1/3+\alpha}$, there exists an explicit function $\qext:\{0,1\}^{n}\to \{0,1\}^m$ such that for any $\ac$ samplable source $X$ over $\F_2$ with entropy at least $k$, it holds that 
\[
\qext(X)\approx_{n^{-\Omega(1)}}U_m
\]
where $m=k^{\Omega(1)}$.
\end{corollary}
\paragraph*{AI Disclosure}
We used an explicit construction of a rank combiner (\Cref{sec:rank_combiner_large_field,sec:field_size_reduction,sec:rank_combiner_final}) provided to us by ChatGPT Astra, after we defined the object. This was one of several ingredients in our proof. We humans did everything else in this paper, including the writing in \Cref{sec:rank_combiner_large_field,sec:field_size_reduction,sec:rank_combiner_final}.
\section{Notation}
We will use capitalized letters to denote random variables. We use the notation $U_t$ to denote a uniform random string on $t$ bits. For a field $\F$, we denote $\F[t]$ to be the polynomial ring over $\F$ and $\F(t)$ to be the field of rational functions over $\F$. We use $\mathbb{N}$ to denote the natural numbers. We use $|X|$ to denote the length of the string $X$. For a function $f$ with domain $\F_2^{n}$, we define $f|_S$ to be the restriction of $f$ to a subset $S\subseteq \F_2^{n}$. We use $\dtv(X,Y)$ to denote the total variation distance (also called statistical distance) between two random variables. 
 For distributions $X,Y$, we say that $X\approx_{\epsilon}Y$ if $\dtv(X,Y)\leq \epsilon$. For random variables $R,S,T$ satisfying $(R,T)\approx_{\epsilon} (S,T)$, we say that $(R\approx_{\epsilon}S)\mid T$. We call a function $f$ a $d$-junta is if it only depends on at most $d$ of its input bits. 

We use $X^{[t]}$ to denote a sequence of random variables $\{X^i\}_{i\in [t]}$.
For a string $x\in \bcube{n}$, $x_{[a,b]}$ denotes the substring of $x$ with starting index $a$ and ending index $b$. When $a=1$, we use the shorthand $x_{[b]}$. 

Given a distribution $X$ on $\F_2^{n}$, for any set $S\subseteq [n]$, we define $X_S\coloneq \{X_i\}_{i\in S}$.  We say that a source $X$ is \textit{flat} if $X$ is uniform over a set. For a set $S$, we define $U_S$ as the random variable that is uniform over it. We use the shorthand $U_m$ to denote a uniform string on $m$ bits. We denote the support of $X$ by $\supp(X)$. Given a set $S\subseteq [n]$, we define the function $\chi_S:\bcube{n}\to \{-1,1\}$ as $\chi_S(x)=(-1)^{\sum_{i\in S}x_i}$.
A source $X$ has $\epsilon$-smooth min entropy $k$, denoted by $\hinf^{\epsilon}(X)\geq k$ if there exists some source $\tilde{X}$ with $\hinf(\tilde{X})\geq k$ such that $X\approx_{\epsilon} \tilde{X}$. A set $S\subseteq \mathbb{N}$ is a Sidon set if $|S+S|\geq \binom{|S|}{2}+|S|$ where $S+S=\{a+b\mid a\in S, b\in S\}$.
\section{Proof Overview}
\label{sec:technical_overview}
Throughout this section we use $C$ and $c$ to denote sufficiently large and small positive constants, respectively.
\subsection{A First Attempt: Black Box Reductions to Affine/Sumset Sources}
\label{sec:barriers}
A natural first attempt at this problem is the following: try to find a sufficiently large affine subspace/sumset in the support of the quadratic source, and then appeal to excellent existing constructions of low entropy affine/sumset extractors. This strategy had already achieved varying levels of success for many other types of sources, including $\nc$ and $\ac$-sources with entropy $C\sqrt{n}$ \cite{viola_circuit_sources,li2016improved}, and small space sources with entropy $C\log n$ \cite{chattopadhyay2022improved,chattopadhyay2022extractors}. 

Unfortunately, there is a fundamental barrier to this approach \cite{alrabiah_et_al:LIPIcs.ICALP.2022.10,polysourceextcgg,alrabiah2024low}. Quadratic maps (and even \(2\)-local maps) can sample \textit{Sidon sets}, which are sets that achieve maximum doubling $(|S+S|=\binom{|S|}{2}+1)$. These are maximally non-affine in a sense, as affine subspaces do not grow under addition. More formally, Sidon sets do not contain affine spaces of dimension \(2\) (see \cite[Claim 5.6]{polysourceextcgg}). This means that any model that can sample (large) Sidon sets cannot be reduced, in a black-box way, to affine sources. In fact, the same holds for sumset sources.

Our approach begins with the questions: Can we exploit the additive structure, or the lack thereof, that makes these reductions fail? Is there a win-win approach to constructing quadratic extractors?

\subsection{A Simple Affine Extractor}
\label{sec:affine}
The above impossibility results show that a generic sumset extractor cannot work. To get around this, we attempt to analyze a specific construction, and see where things break down. We will then exploit the structure from the barrier results to overcome these failures. 

A central object in almost all known affine extractors is a \textit{strong linear seeded extractor}. 

\begin{definition}[Strong Linear Seeded Extractor]
\label{defn:strong_linear_seeded_extractor}
  A function $\lext:\cube{n}\times \cube{d}\to \cube{m}$ is a $(k,\epsilon,\delta)$-strong linear seeded extractor with seed length $d$ and output length $m$ if 
  \begin{enumerate}
      \item the function $\lext_{y}(x)=\lext(x,y)$ is linear as a function of $x$ for all seeds $y$, and 
      \item for every source $X$ with $\hinf(X)\geq k$, with probability at least $1-\epsilon$ over $y\sim \cube{d}$, it holds that $\lext(X,y)\approx_{\delta}U_m$.
  \end{enumerate}
  When $\epsilon=\delta$, we refer to the function as a $(k,\epsilon)$-strong linear seeded extractor. If item~2 above only works for sources from a class $\mathcal{X}$, then we say that the function is an $(\epsilon,\delta)$-strong linear seeded extractor for $\mathcal{X}$. We refer to $\epsilon$ as the strength error and $\delta$ as the closeness error. 
\end{definition}

We start by analyzing one of the simplest affine extractors:\footnote{There are simpler affine extractors (like inner product), but the following extractor will be most relevant to our construction.}
\[
\ext(X)=\lext(X,X_{[d]})
\]
where $X_{[d]}$ outputs the first $d$ bits in $X$ and $d$ is appropriately chosen. 

We first recall the proof of why $\ext(X)=\lext(X,X_{[0.1n]})$ is an affine extractor for affine sources with entropy $0.99n$. We choose $\lext$ to be a $(0.1 n,\epsilon)$-strong linear seeded extractor with seed-length $0.1n$, outputting $1$ bit with $\epsilon=2^{-0.05n}$ (we can use for instance an extractor from \cite{chattopadhyay2021affine}). 
The proof involves two main steps that we now discuss.
\paragraph{Affine Conditioning} The first step in the proof is the following lemma from \cite{rao2009extractors}, known as \textit{affine conditioning}.
\begin{lemma}[Affine Conditioning, \cite{rao2009extractors}]
\label{lem:affine_conditioning}
    Let $X$ be an affine source with $\hinf(X)\geq k$ and suppose $L:\F_2^{n}\to \F_2^{m}$ is a linear map. Then, $X=A+B$ for independent affine sources $A,B$ such that \begin{enumerate}
    \item $\hinf(A)\geq k-m$, 
    \item $\hinf(B)\leq m$, and 
    \item $L(a)=0$ for all $a\in \supp(A)$.
    \end{enumerate}
\end{lemma}
Before explaining why this lemma is useful, we briefly sketch its proof. Observe that an affine source with entropy $k$ can be represented as $X=v+M(U_k)$ for a full rank linear map $M$ and fixed vector $v$. Now, the map $L\circ M$ has rank at most $m$; hence its null space $P$ has dimension at least $k-m$. Hence, $\F_2^{k}=P+Q$ for $Q$ with dimension at most $m$. Taking $A=M(P)$ and $B=v+M(Q)$ completes the proof. 

Now, going back to our candidate extractor $\ext$, observe that the function $S(X)=X_{[0.1n]}$ is linear. Thus, we have 
\begin{equation}
    \label{eqn:affine_simple_ext_1}
\ext(X)=\lext(X,X_{[0.1n]})=\lext(A+B,S(A+B))=\lext(A+B,S(B))
\end{equation}
where $A$ and $B$ are the independent affine sources given by \Cref{lem:affine_conditioning} satisfying (1) $\hinf(A)\geq 0.89n$ and (2) $\hinf(B)\leq 0.1 n$. 

\paragraph{Using Linearity} From the linearity of $\lext$ after fixing the seed, we have that $\ext(X)=\lext(A,S(B))+\lext(B,S(B))$. 
Suppose first that $S(B)$ is uniform. The guarantee of the strong seeded extractor would imply that $\lext(A,S(b))$ is close to uniform with high probability over fixing $B=b$. The second term becomes constant after fixing $B$ and thus does not affect the uniformity. 

Now, we remove this restriction and consider our scenario where $S(X)$ is a prefix of $X$ of length $0.1 n$. The fact that $\hinf(X)\geq 0.99n$ implies that $\hinf(S(B))\geq 0.09 n$. This implies that the probability of $S(B)$ being in any set of seeds is at most $2^{0.01n}$ times the corresponding probability over the uniform distribution. Our choice of $\epsilon=2^{-0.05 n}$ implies that $S(B)$ is a good seed with high probability and completes the proof. 

\subsection{Extending to Quadratics}
\label{sec:extend_quadratics}
 We will now try to analyze the previous construction for quadratic sources, but we will see that a few problems arise.  We take $\ext(X)=\lext(X,X_{[d]})$, where we now choose $d=n^c$ for some small constant $c>0$. We are again targeting the near-linear entropy regime $k\geq n-O(n^c)$ (note that the previous best known extractors from \cite{polysourceextcgg} required entropy $n-O(\sqrt{\log n}))$.  Throughout most of this discussion, we assume that our quadratic source $X=f(U_k)$ satisfies $\hinf(X)=k$. This is not without loss of generality: unlike affine sources, a quadratic source need not admit a representation where the number of generating bits matches the entropy. We make this assumption for simplicity, and in our actual proof, we use an input reduction lemma to remove it (\Cref{lem:input_reduction_entropy}). 
\paragraph{Quadratic Conditioning}
The first part of our argument will generalize the affine conditioning step in the previous argument. Clearly, the lack of linearity in our source prevents us from using the rank based argument from earlier. Regardless, we will argue the following.
\begin{lemma}[Informal]
\label{lem:quadratic_conditioning_technical_overview}
    Let $X=f(U_k)$ be a quadratic source and let $p:\F_2^{k}\to\F_2^{m}$ be a quadratic map. Then $(f(U_k),p(U_k))$ is $\epsilon$-close to a convex combination of sources of the form $(f(A+B),q(B))$ where $A$ and $B$ are independent affine sources in $\F_2^{k}$ such that \begin{enumerate}
        \item $\hinf(A)= \Omega(k/m)$.
        \item $\hinf(B)=\ell\leq m+O(\log(1/\epsilon))$,
        \item $q(B)=p(a+B)$ for all $a\in \supp(A)$,
        \item $q(B)\approx_{\epsilon} p(U_k)$, and 
    \end{enumerate}
\end{lemma}

The lemma states that the source can be decomposed into sources satisfying the property that $p$ only depends on a low dimensional affine subspace (property 2 and 3) of the generating bits. Furthermore, this decomposition preserves the distribution of $p$ (property 4). The analogy in the affine case is taking the linear function $\ell=L\circ M$ (where $M$ is the matrix generating the source). It is also crucial that there is entropy in the source independent of $p$. This is ensured by property 1 (note the multiplicative loss this time, compared to the additive loss in the affine case).
We now sketch a proof of \Cref{lem:quadratic_conditioning_technical_overview}. 
\begin{proof}[Proof Sketch of \Cref{lem:quadratic_conditioning_technical_overview}]
There are two main steps. First, we argue that the distribution of $p$ is preserved even after fixing all but $m+O(\log(1/\epsilon)$ generating bits. We then argue that we can do this fixing efficiently, where we make $p$ independent of the rest of the generating bits without trivially fixing all of them.

The key tool for the first step is the \textit{input reduction lemma} due to Chattopadhyay, Goodman and Gurumukhani \cite{polysourceextcgg}.
\begin{lemma}[Input Reduction: Short Output Length \cite{polysourceextcgg}]
    \label{lem:input_reduction_short}
    Let $p:\bcube{k}\to \bcube{m}$ be a quadratic map. For any $0<\epsilon<1/4$, for a random linear map $L:\bcube{k}\to \bcube{k-\ell}$ with $\ell=m+O(\log(1/\epsilon))$ it holds with probability at least $1-\epsilon$ that
    \[
    (p(U_k),L(U_k))\approx_{\epsilon}(p(U_k),U'_{k-\ell}))\] where $U', U$ are independent, uniform random variables.
\end{lemma}
      
    Let $e_1,\ldots, e_\ell$ be a basis for the nullspace of $L$ and $e_{\ell+1},\ldots, e_{k}$ be a extension of the basis to all of $\F_2^{k}$. We have that $U_{k}=A+B$ for independent $A,B$ where  $A$ is uniform over the span of $\{e_{\ell+1}, \ldots, e_{k}\}$ and $B$ is uniform over the nullspace of $L$. We can rewrite the polynomial $p$ as being a quadratic function of $A,B$. We have 
    \[
    p(A+B)=p_1(A)+p_2(B)+p_3(A,B)
    \] where $p_3$ is bilinear. 
    From \Cref{lem:input_reduction_short}, we now have that with probability $1-\epsilon$ over $a\sim A$, it holds that $p(a+B)\approx_{{\epsilon}}p(U_k)$. This completes the proof of the first step.

    We now implement the second step. We will convert $p$ to a function $q$ solely depending on $B$ by fixing all dependencies on $A$. We want to do so while still retaining entropy in $A$. Note that $p_3$ depends on $m\ell$ linear functions of $A$. We will fix these $m\ell$ linear functions along with $p_1(A)$ which is a quadratic function. However, this causes an issue when compared to the affine case: $A$ becomes a quadratic variety after fixing $p_1(A)$. The additional guarantee that $p_1$ only has $m$ output bits now comes to the rescue. We use the following theorem by Cohen and Tal \cite{cohen-tal}.\footnote{The use of Cohen-Tal to retain affine structure after fixing a few quadratics (in the context of extractors) appeared in \cite{DG26}.}
\begin{theorem}[{\cite[Theorem 3.6]{cohen-tal}}]
    \label{thm:cohen-tal}
    Let $q_1,\ldots,q_t$ be quadratic functions from $\bcube{k}\to\F_2$. There exists a universal constant $C$ (independent of $q_1,\ldots,q_t$) such that $\bcube{k}$ can be partitioned into affine subspaces each of dimension $d=C(k/t)-t$ where $q_1,\ldots,q_t$ are constant when restricted to each of these subspaces.
\end{theorem}
The above theorem is saying that the input space of a small set of quadratic polynomials can be partitioned with affine subspaces such that these polynomials are constant on each of these subspaces. This is exactly what we need to fix $p_1(A)$. \Cref{thm:cohen-tal} allows us to fix $p_1(A)$ by  breaking $A$ into affine subspaces where $p_1$ is constant (we need to also fix $m\ell$ linear functions of $A$ to make $p_3$ a function of $B$). Each such subspace $A'$ has dimension $\Omega(k/m)$ as long as $k\gg m\ell$, which will be the case in all our applications. Renaming $A'$ as $A$ completes the proof.
\end{proof}

Since $A$ and $B$ are affine subspaces of $\F_2^{k}$, we can rewrite $f(A+B)=f'(Y,Z)$ where $Y=U_{\dim(A)}$, $Z=U_{\dim(B)}$, and $f'$ is a quadratic map. Similarly, $q(B)=q'(Z)$ for some quadratic $q'$. 

For our running example $\ext(X)=\lext(X,X_{[d]})$, we take $p=(f(U_k))_{[d]}$ with $d=n^c$ as the quadratic in \Cref{lem:quadratic_conditioning_technical_overview}. Assuming $k\geq n-0.01 n^c$, we have $\hinf(p(U_k))\geq 0.99d$. Thus, finally after applying the decomposition from \Cref{lem:quadratic_conditioning_technical_overview}, and using the renaming above, it suffices to analyze random variables of the form $\lext(f(Y,Z),q(Z))$ where $Y=U_{s}$ for $s=\Omega(k/d)$ and $\hinf^{\epsilon}(q(Z))\geq 0.99 d$.  
\paragraph{Obtaining Linearity via Cauchy Schwarz}
Similar to the affine case, we analyze the case where $q(Z)=U_{d}$ is exactly uniform (the $0.99d$ entropy case can be easily handled by using the same argument as in \Cref{sec:affine} and we omit it here).

Thus, our setup is the following: we have a source $X=f(Y,Z)$ with $Y=U_s$ for $s=\Omega(k/d)= \Omega(n^{1-c})$ and $q(Z)=U_{n^c}$. We are analyzing the function $\lext(X,q(Z))=\lext(f(Y,Z),q(Z))$.

It suffices to argue that all the parities are unbiased, as we can infer near-uniformity from Vazirani's XOR lemma. The affine argument clearly does not extend: $f$ is a quadratic and hence we can't break it into two terms where the first is independent of $Z$.  Let $b(z)=|\E_{Y}[\chi_S(\lext(f(Y,z),q(z)))]|$ and $f(Y,Z)=f_1(Y)+f_2(Y,Z)+f_3(Z)$ where $f_2$ is bilinear. Our insight is to use the Cauchy-Schwarz inequality. This was inspired by other applications of Cauchy-Schwarz in the proofs of correlations bounds, xor lemmas and pseudorandom generators for $\F_2$-polynomials \cite{bns92,bogdanov_viola,lovett_polynomials_prgs,viola_prg_polynomial,viola_wigderson}. We have that 
        \begin{align*}
            \E_{Z}[(b(Z))^2]&=\E_{Z}(\E_{Y}[\chi_S(\lext(f(Y,Z),q(Z)))])^2\\
            &=\E_{Y,Z}\E_{Y,A}[\chi_S(\lext(f(Y,Z)+f(Y+A,Z),q(Z)))]\\
            &=\E_{Z}\E_{Y,A}[\chi_S(\lext(\partial_A f_1(Y)+f_2(A,Z),q(Z)))
        \end{align*}
        where $\partial_a g(x)\coloneq g(x)+g(x+a)$ is the direction derivative in the direction $a$.

        Observe that for any fixed $a$, we have that $\partial_a f_1(.)$ is an affine source! Also, $f_2(a,Z)$ is only a function of~$Z$. Now, we can use linearity similar to the affine proof. We have 
        
        \begin{equation*}
            \E_{Z}[(b(Z))^2]\leq2^{-s}+\max_{a\neq 0}\E_{Z}|\E_{Y}[\chi_S(\lext(\partial_a f_1(Y),q(Z)))|.
        \end{equation*}

        We have reduced the problem to bounding the error of a linear seeded extractor when applied to the derivative sources $\partial_a f_1(Y)$. A sufficient condition for the error to be small is if $\min_{a\neq 0}\hinf(\partial_af_1(Y))$ is large (greater than the entropy requirement of $\lext$). We call these quantities "derivative entropies". We record this argument in the following lemma. 
        
\begin{lemma}[Informal, see \Cref{thm:cauchy_schwarz_general}]
    \label{thm:cauchy_schwarz_general_overview}
        Let $X=f(Y,Z)$ be a source with $f$ quadratic, $Y=U_{s}$, $Z=U_t$ independent of $Y$ and such that $f(Y,Z)=f_1(Y)+f_2(Y,Z)+f_3(Z)$ where (1) $\min_{a\neq 0}\hinf(\partial_a f_1(Y))\geq k$ and (2) $f_2$ is bilinear. Let $\lext:\bcube{n}\times \bcube{d}\to \bcube{m}$ be a $(k,\epsilon)$ strong linear seeded extractor and $q(Z)$ be a function of $Z$ satisfying $q=U_d$. Then, it holds with probability $1-2^{m}\cdot (2^{-s}+2\epsilon)^{1/2}$ over $z\sim U_t$ that 
        \[
        \dtv(\lext(f(Y,z),q(z)),U_{m})\leq 2^{m/2}(2^{-s}+2\epsilon)^{1/4}.
        \]
    \end{lemma}
        
        It is still unclear if we have made any progress, as we do not know any bounds on these derivative entropies.

        Now, recall our discussion on the barriers to reducing the whole problem to affine sources in \Cref{sec:barriers}. The bad examples were Sidon sets: sets which have maximum doubling. Remarkably, these are precisely the good cases for our discussion in this section: Sidon sets have maximum derivative entropy! Indeed, for a source representing a Sidon set, say $X=f(U_k)$, we have that for any $a\neq 0$, $f(x)+f(x+a)$ is distinct for every unordered pair $(x,x+a)$, which implies $\hinf(\partial_a f(U_k))=k-1$. This is the starting point of our win-win analysis. 
\subsection{Structure Theorem for Quadratics}
\label{sec:structure_theorem_overview}
The main ingredient for our win-win analysis is a structural result on quadratic sources. 
\begin{theorem}
\label{thm:structure_injective}
    Let $X$ be a quadratic source with an injective generating map $f:\mathbb{F}^{k}_2\to \mathbb{F}^{n}_2$. Let $r,t\in \mathbb{N}$ be such that $rt\leq k/2$. Then, $X$ can be exactly expressed as a convex combination of sources of the following two types: 
    \begin{enumerate}
        \item Affine sources with entropy at least $t$ (obtained by applying $f$ to an affine subspace $V$ such that $f|_V$ is affine), or 
        \item Quadratic sources with an injective generating map $g:\bcube{k'}\to \bcube{n}$ satisfying (1) $\min_{a\neq 0}\hinf(\partial_a g(U_{k'}))\geq r$, (2) $k'\geq k-rt$ and (3) $g=f\circ L$ for an injective affine map $L:\F_2^{k'}\to\F_2^{k}$. 
    \end{enumerate}
\end{theorem}
Again, for simplicity we assume that the number of generating bits matches the entropy. The above theorem says that any quadratic source is a convex combination of (1) high entropy affine sources, or (2) sources with high derivative entropy. As discussed in the previous sections, this dichotomy seems useful, as we could try employing an affine analysis for the first case, and a quadratic analysis (using quadratic conditioning and  Cauchy-Schwarz) for the second. We now give a proof sketch. 
\begin{proof}[Proof Sketch of \Cref{thm:structure_injective}]
Our proof of the theorem relies on a simple observation. Consider a quadratic source $X=f(U_k)$ with $\partial_a f(.)=c$ for some non-zero direction $a$ and constant vector $c$. This implies that the quadratic function looks linear when viewed along the direction $a$. Formally, take a basis $B$ of dimension $k-1$ such that $\F_2^{k}=\vecspan\{\{a\}\cup B\}$. We have that 
\[
f(ya+b)=y\partial_a f(b) +f(b)=yc+f(b)
\]
for any bit $y$ and $b\in B$. Thus taking a random bit $Y$ and a random variable $Z$ uniform over $B$ and independent of $Y$, we have that original source is can be expressed as
\[
g(Y,Z)=Yc+f(Z).
\]
Thus, we have obtained a one dimensional affine source from our original source $f$:   fixing $Z$ decomposes $X$ into a convex combination of shifts of these one dimensional affine sources.

The idea is now to do this linearizing procedure every time we find a direction $a$ with $\hinf(\partial_{a}f(.))\leq r$. For any such direction, we condition on the value of $\partial_a f(.)=c$. This loses at most $r$ bits of entropy by our definition of $a$. Now after conditioning, we can use the linearization trick from above to obtain a one dimensional affine source. We repeat this $t$ times as long as $rt\leq k$ to obtain the decomposition stated in the theorem statement.
\end{proof}

We now define the notion of a "pure" quadratic source: a quadratic source that does not contain a large dimensional affine subsource. Our analysis will take different paths based on whether the source is pure or not. 
\begin{definition}[Pure Quadratic Source]
\label{def:pure_quadratic}
    A quadratic source $X=f(U_s)$ is $q$-pure if there exists no affine subspace $V\subseteq \F_2^{s}$ such that $f|_{V}$ is affine and $\dim(f(V))\geq q$.
\end{definition}

We note that the above notion of \textit{purity} depends on both the source $X$ and the generating map $f$. This definition of pure quadratic gives us a natural corollary of \Cref{thm:structure_injective}.

\begin{corollary}[Pure Quadratics can be decomposed into high derivative entropy sources]
\label{clry:pure_quadratic_structure_informal}
      Let $X=f(U_k)$ be a $t$-pure quadratic source with $\hinf(X)=k$. Then, $X$ can be expressed as a convex combination of sources of the form $X'=g(U_{k'})$ satisfying  (1) $\min_{a\neq 0} \hinf(\partial_a g(.))\geq r$ and (2) $k'\geq k-rt$.
\end{corollary}
\subsection{Warmup: Extractor for Pure Quadratics}
Recall that our goal was to analyze the construction $\ext(X)=\lext(X,X_{[n^c]})$ for quadratic sources with min-entropy greater than $n-n^c$ for small constant $c$. 

In \Cref{sec:extend_quadratics}, we showed via \textit{quadratic conditioning} that it suffices to study random variables of the form 
\[
\lext(f(Y,Z),q(Z))
\]
where $X=f(Y,Z)$ with $Y=U_s$ for $s\geq  \Omega(n^{1-c})$ and $\hinf^{\epsilon} (q(Z))\geq 0.99n^c$. For simplicity, we assume $q(Z)=U_{n^c}$ (the high entropy version can be handled similar to \Cref{sec:affine}).

Let $f(Y,Z)=f_1(Y)+f_2(Z)+f_3(Y,Z)$ for bilinear $f_3$. In \Cref{thm:cauchy_schwarz_general_overview}, we established  near-uniformity of $\lext(f(Y,Z),q(Z))$ assuming $\min_{a\neq 0}\hinf(\partial_a f_1(.))$ is large. We are now ready to use our techniques from \Cref{sec:structure_theorem_overview} to obtain this structure when the source is pure. We use a slightly stronger version of \Cref{clry:pure_quadratic_structure_informal}.

\begin{lemma}[Informal, see \Cref{clry:pure_quadratic_structure}]
\label{clry:pure_quadratic_structure_informal_overview_pure_extractor}
      Let $X=f(Y,Z)$ be a $t$-pure quadratic source with $\hinf(X)=k$. Then, for any $rt\leq s/2$, $X$ is exactly expressed as a convex combination of sources of the form 
      \[
      X'=g(Y',Z)=g_1(Y')+g_2(Y',Z)+g_3(Z)
      \]
      such that (1) $\min_{a\neq 0}\hinf(\partial_a g_1(Y'))\geq r$, (2) $\hinf(X')\geq s'+|Z|$, (3) $Y'=U_{s'}$ with $s'\geq s/2$, and (4) $g_2$ is bilinear. Importantly, the random variable $Z$ is left unchanged by the decomposition.
\end{lemma}
The above theorem is proved similar to \Cref{thm:structure_injective}, except that we only fix functions of $Y$ in the proof and leave $Z$ unchanged.

Note that the first property is exactly what we wanted for \Cref{thm:cauchy_schwarz_general_overview}. Now choosing $r$ greater than the entropy requirement of $\lext$ (say $r=n^{0.5}$) implies the following theorem.
\begin{theorem}[Informal]
    Let $X$ be an $n^{0.2}$-pure quadratic source with $\hinf(X)\geq n-n^c$ for a small enough constant~$c>0$. Then $\ext(X)=\lext(X,X_{[n^c]})\approx U_m$.
\end{theorem}

This argument works well when the source is a pure quadratic (for a formal proof, see \Cref{lem:A is uniform}). Unfortunately, there is an issue when the source is not pure (contains a large subspace). Note that for any useful bound on derivative entropy, we need to choose $t\leq s/(2r)=o(n)$ in \Cref{clry:pure_quadratic_structure_informal_overview_pure_extractor}. Thus, when the source is not pure, we are only guaranteed an affine subspace of entropy $o(n)$. However, this entropy is too small for the candidate construction ($\ext$ is only an affine extractor for entropy $0.99n$ or so, see \Cref{sec:affine}). 

If we are happy with a 2-bit disperser (whose output is non-constant), then there is a simple construction: Output $(\ext(X),\affinext(X))$ where $\affinext(X)$ is a low entropy affine extractor (say from \cite{li2016improved}). When the source is pure, the first bit is uniform, and when the source is not pure, the second bit is non-constant.

It turns out that with one more idea, we can actually use this approach to construct a 1-bit disperser (for entropy $n-n^{0.1}$, which already surpasses prior work). We use the Challenge-Response framework from \cite{BKSSW10} which has proven very successful in constructing two source and affine dispersers \cite{beating_frankl_wilson,shaltiel_affine,cohen2016two}. We use this framework to detect if we are in the pure quadratic or affine case with a single sample. We find this application of the Challenge-Response framework quite interesting, as it different from its usual usage. We defer this construction to the appendix (see \Cref{thm:main_disp}). 
\subsection{Low Entropy Extraction: the CZ Framework and  Li's Affine Extractor}
We now begin our discussion of our construction for the polylogarithmic entropy regime. Our construction follows the framework introduced by Chattopadhyay and Zuckerman \cite{CZ19} to construct polylogarithmic entropy two source extractors. We refer the reader to the wonderful survey by Chattopadhyay \cite{eshan_survey} for an exposition that starts from first principles. This framework was applied to affine sources by Li \cite{li2016improved}. Analogous to the previous sections, we will consider this affine extractor and then see what has to be adapted for quadratic sources. 
\paragraph{Overview of CZ construction and $(q,t)$-NOBF sources}
At a high level, the CZ framework, including Li's construction, has two steps: (1) reduce the source $X$ into a $(q,t)$-non oblivious bit fixing (NOBF) source, and (2) apply an explicit extractor for these sources. 

To make sense of the above description, we need to define the notion of a $(q,t)$-NOBF source.
\begin{definition}[$(q,t)$-non oblivious bit fixing (NOBF) sources]
    A distribution $X$ on $\F_2^{n}$ is a $(q,t)$-NOBF source if there exists a set $Q$ of size at most $q$ such that $X_{[n]\setminus Q}$ is $t$-wise independent. 
\end{definition}
Informally, these sources have the following property: outside $q$ adversarial bits, the rest of the distribution is $t$-wise independent. One of the main technical achievements in \cite{CZ19} was the construction of explicit extractors (with a single output bit) for these sources (also known as \textit{resilient functions}\footnote{technically, resilient functions only need to work for $(q,n-q)$-NOBF sources}). Subsequently, there were improvements by Li \cite{li2016improved} (obtaining multiple output bits) and Meka \cite{meka2017explicit} (obtaining almost optimal parameters matching the non-explicit bound of \cite{ajtai1993influence}).

For our purposes, we use the following theorem.
\begin{theorem}[Explicit NOBF Extractor \cite{CZ19,li2016improved}]
\label{thm:nobf_extractor}
    For any positive constant $\delta\in (0,1)$, there exists an explicit function $\nobfext:\{0,1\}^{n}\to \{0,1\}^m$ and constant $C$ such that for any $(n^{1-\delta},(m\log n)^C)$-NOBF source $X$, it holds that $\nobfext(X)\approx_{n^{-\Omega(1)}} U_m$.
\end{theorem}

\paragraph{Reduction to $(q,t)$-NOBF sources: }
Given the above NOBF-extractor, all that is left to do is convert the original affine source $X$ into a $(q,t)$-NOBF source for $q=N^{1-\delta}$ and $t=\log^C N$ for some $N$ that we choose. Li \cite{li2016improved} uses two components to achieve this: (1) a \textit{strong linear seeded extractor} and (2) an \textit{affine correlation breaker}.

We have already seen the definition of the first object \Cref{defn:strong_linear_seeded_extractor}. We now define the second.
\begin{definition}[Affine Correlation Breaker]\footnote{The standard definition of an affine correlation breaker is more general, we state it this way for ease of exposition.}
    \label{defn:affine_correlation_breaker}
A function $\acb:\F_2^{n}\times \F_2^{d}\times \cube{a}\to \cube{m}$ is a $(t,k,\delta)$-affine correlation breaker if it holds that for random variables $X,W,W^{[t]}$ and advice strings $\alpha,\{\alpha^i\}_{i\in [t]}$ in $\cube{a}$ satisfying
\begin{enumerate}
    \item $X=f(V)$ is an affine source with $\hinf(X)\geq k$ and $V=U_k$,
    \item $W=\ell(V)$, $W^{i}=\ell^{i}(V)$ for all $i\in [t]$ with $\ell,\ell^1,\ldots,\ell^t:\F_2^{k}\to\F_2^{d}$ are affine maps, 
    \item $W=U_d$,
    \item $\alpha\neq \alpha_i$ for all $i\in [t]$, 
\end{enumerate}
it holds that 
\begin{equation*}
    (\acb(X,W,\alpha)\approx_{\delta} U_m)\mid \{\acb(X,W^{i},\alpha^{i})\}_{i\in [t]}
\end{equation*}where $U_m$ is uniform random string independent of $X,W^{[t]}$.
\end{definition}
Li \cite{li2016improved} gives the first explicit construction of this object and directly uses it to reduce affine sources to $(q,t)$-NOBF sources. His reduction is the following: 
\begin{enumerate}
    \item Compute $W=\{\lext(X,i)\}_{i\in [N]}$
    \item Output $S=\{\acb(X,W^i,i)\}_{i\in [N]}$.
\end{enumerate}
Here $\lext$ is a strong linear seeded extractor with seedlength $\log N=O(\log n)$ (we treat the seeds as integers). For an affine source, one can show that all but $q=N^{1-\delta}$ of the $(W^i)$'s are truly uniform. Now, given any set of $t$ good indices, the definition of $\acb$ implies that $S_T$ is almost $t$-wise independent with error $\delta t$. Assume $\delta$ is sufficiently small, one can argue that $S$ is close to a $(q,t)$-NOBF source \cite{agm03}. Li's construction then outputs $\nobfext(S)$ and our argument implies that it is an affine extractor.
\subsection{Quadratic Correlation Breaking}
One immediate issue when extending to quadratic sources $X=f(U_k)$ is that the random variables $W^i$ output by the linear seeded extractor are now quadratic functions of the generating bits. These are not handled by the definition of the affine correlation breaker. This motivates the definition of a \textit{quadratic correlation breaker}. 
\begin{definition}[Quadratic Correlation Breaker, see \Cref{defn:quadratic_correlation_breaker} for a more precise version]
    \label{defn:quadratic_correlation_breaker_overview}
A function $\qcb:\F_2^{n}\times \F_2^{d}\times \cube{a}\to \cube{m}$ is a $(t,k,q,\delta)$-quadratic correlation breaker if it holds that for random variables $X,W,W^{[t]}$ and advice strings $\alpha,\{\alpha^i\}_{i\in [t]}$ in $\cube{a}$ satisfying
\begin{enumerate}
    \item $X=f(V)$ is a $q$-pure quadratic source with $\hinf(X)\geq k$ and $V=U_k$,
    \item $W=p(V)$, $W^{i}=p^{i}(V)$ for all $i\in [t]$ with $p,p^1,\ldots,p^t:\F_2^{k}\to\F_2^{d}$ are quadratic maps, 
    \item $W=U_d$,
    \item $\alpha\neq \alpha_i$ for all $i\in [t]$, 
\end{enumerate}
it holds that 
\begin{equation}
    \label{eqn:qcb_guarantee_main_overview}
    (\qcb(X,W,\alpha)\approx_{\delta} U_m)\mid \{\qcb(X,W^{i},\alpha^{i})\}_{i\in [t]}
\end{equation}where $U_m$ is uniform random string independent of $X,W^{[t]}$.
\end{definition}

Note that our definition of quadratic correlation breaker requires the underlying source to be a pure quadratic source. This will be crucial for our construction. We show that an existing affine correlation breaker from \cite{chattopadhyay2022extractors} also satisfies the quadratic correlation breaking guarantee (see \Cref{thm:quad_correlation_breaking}). This allows us to execute the win-win analysis: if the source is not-pure, we employ the affine correlation breaker analysis, else we use our new quadratic analysis.

Our proof of \Cref{thm:quad_correlation_breaking} uses all the tools we developed in the previous sections: quadratic conditioning, Cauchy-Schwarz and the structure theorem. We refer the reader to \Cref{sec:quadratic_correlation_breaking} for the full proof. We now sketch an argument, assuming familiarity with standard advice correlation breakers (see \Cref{defn:correlation_breaker}).

The construction we use, proposed by \cite{chattopadhyay2022extractors} uses a standard advice correlation breaker $\CB$ and a strong linear seeded extractor $\lext$ as black-box objects. We define $\qcb(X,W,\alpha)$ in the following way:
\begin{enumerate}
    \item $S_1=W_{[d_0]}$
    \item $R=\lext(X,S_1)$
    \item Output $S=\CB(W,R,\alpha)$.
\end{enumerate}
Here $d_0\ll d$ is the seedlength of $\lext$ and the output length of $\lext$ matches the seedlength of $\CB$.

We now sketch our quadratic analysis that works for pure quadratics. Let $S_1^{[t]}$, $R^{[t]}$ and $S^{[t]}$ be the tampered versions. Our goal is to argue that $S\approx U_m\mid S^{[t]}$. In the first step, we argue that $R$ is uniform, even conditioned on $S_1$. To do, we first apply quadratic conditioning (from \Cref{lem:quadratic_conditioning_technical_overview}) with respect to $\{W,W^{[t]}\})$ and reduce to analyzing $X=f(Y,Z)$ and $W=p(Z),W^{[t]}=p^{[t]}(Z)$ ($Y,Z$ are still uniform, but over fewer bits). Now, we also have that $S_1,S_1^{[t]}$ are solely functions of $Z$. 

We now use the fact that $X$ is a pure quadratic. We apply the structure theorem \Cref{clry:pure_quadratic_structure_informal_overview_pure_extractor} and the Cauchy Schwarz argument \Cref{thm:cauchy_schwarz_general_overview} to conclude that $R$ is uniform (even conditioned on $Z,S_1$).

Now, our goal is to make the sources of the form required by the final advice correlation breaker. For this, we need to make $W,W^{[t]}$ independent of $R,R^{[t]}$. Our goal is to do this while maintaing the uniformity of $R$ (as $\CB$ requires a uniform seed). It turns out the analogous step in affine case is relatively straightforward. For us, achieving this independence needs us to again use the input reduction lemma (now with respect to $R$) and carefully fix information so that $R$ remains uniform and $W$'s entropy is not fully destroyed (this is important as $\CB$ needs a high entropy source). We skip this argument here and refer to the proof of \Cref{thm:intermediate_qcb} (we use a stronger version of the input reduction lemma, see \Cref{lem:partial_input_reduction}). 

Finally, after $R,R^{[t]}$ and $W,W^{[t]}$ are made independent, $R$ is uniform, and $W$ has entropy, we can appeal to the guarantee of the correlation breaker $\CB$ to complete the proof.
\subsection{Good Linear Seeded Extractors for Quadratic Sources: Rank Combiners}
Recall that Li's construction first computed $W=\{\lext(X,i)\}_{i\in [N]}$ and then passed these through the affine correlation breaker as $S=\{\acb(X,W^{i},i)\}_{i\in [n]}$ to obtain the final $(q,t)$-NOBF source. Now, the natural extension for us would be to use the quadratic correlation breaker that we constructed above in the second step. This seems reasonable as each $W^i$ is now a quadratic polynomial of the generating bits of the source (as they are linear functions of the quadratic source) and $\qcb$ was defined precisely to handle such quadratically correlated seeds.

However, there is a subtle issue. Note that the good entries of $W$ are $\epsilon$-close to uniform, where $\epsilon$ is at most the error of $\lext$. Passing this through the correlation breaker implies that roughly, the good bits (corresponding to near uniform entries in $W$) are almost $t$-wise independent with error $\epsilon t$. The issue is that the result of \cite{agm03} needs $\epsilon$ to be close to $N^{-t}$ for completing the proof and showing that $S$ is close to a $(q,t)$-NOBF source. However, $N\geq 1/\epsilon^2$ is the minimum seed length required for $\lext$ and so such small error is not achievable anymore. This issue was completely avoided in the affine case, as one can show that the near-uniform bits of $W$ obtained from an affine source
are in fact exactly uniform (as linear functions are either constant or have zero-bias).

This is no longer the case for quadratics, as now we can have almost any bias for a non constant quadratic function (take inner product on $2b$ bits to get bias $2^{-b}$). We resolve this issue by using linear seeded extractors with unusually good error on the good seeds (when applied to quadratics with high derivative entropy).

\begin{theorem}[See \Cref{clry:explicit_rank_combiner_extractor} for formal statement]
\label{clry:explicit_rank_combiner_extractor_overview}
For any $k,d\geq \log^C n$, and $\epsilon\geq 1/n^2$, there exists an explicit function $\rcombext:\F_2^{n}\times [N]\to \F_2^{d^{\Omega(1)}}$ with $N=\poly(n)$ such that  
 $\rcombext$ is a $(\epsilon,2^{(-d^{\Omega(1)})})$-strong linear seeded extractor for the class of quadratic sources $X=f(U_k)$ satisfying $\min_{a\neq 0}\hinf(\partial_a f(\cdot))\geq d$. $\rcombext$ is also an $(\epsilon,0)$ strong linear seeded extactor for affine sources with entropy at least $k$.
\end{theorem}
We remind the reader that an $(\epsilon,\delta)$ strong linear seeded extractor is one where at most $\epsilon$ fraction of the seeds have error greater than $\delta$. 
To construct this object, we introduce a new pseudorandom object that we named a \textit{rank combiner}.
\begin{definition}[Rank Combiner, see \Cref{def:rank_combiner} for a more general definition]
\label{def:rank_combiner_overview}
   A $(d,r,\epsilon)$-rank combiner over a field $\mathbb{F}$ is a collection of vectors $w_1,\ldots, w_N$ in $\mathbb{F}^{n}$ such that for all  square matrices $G_1,\ldots, G_n$ satisfying  $\min_{y\neq 0}\dim \{G_iy\}_{i\in [n]}\geq d$, it holds with probability $1-\epsilon$ over $t\sim [N]$ that
    \begin{equation}
    \label{eqn:rank_combiner_overview}
        \rank_{\mathbb{F}}\left(\sum_{i=1}^{n}(w_t)_{i}G_i\right)\geq r
    \end{equation}
\end{definition}
Note that property $\min_{y\neq 0}\dim\{G_iy\}_{i\in [n]}\geq d$ is just a matrix reformulation of the derivative entropy property\footnote{Another reformulation is that the vector space of matrices $\mathcal{G}=\{[G_1 y,\ldots, G_n y]\}_{y}$ is a rank-metric code with distance $d$.}. Thus, constructing rank combiners over $\F_2$  imply \Cref{clry:explicit_rank_combiner_extractor_overview}\footnote{The definition in \Cref{eqn:rank_combiner_overview} corresponds to one output bit. For a version corresponding to multiple output bits, see \Cref{def:rank_combiner}.}. ChatGPT Astra gave us such a construction (\Cref{thm:explicit_rank_combiner_f2}).

The notion of a rank combiner seems interesting in its own right. It adds to the rich body of work in algebraic pseudorandomness, studying objects such as rank extractors \cite{GR08,dvir2009extractors}, rank condensers \cite{FS12,FSS14}, subspace designs \cite{GX13,gk_subspace_designs} and the various connections between them \cite{FG15}.

Rank combiners do not seem reducible to any of these existing objects. However, Astra's proof uses various ideas originating from this line of work. We present the proof and more details on connections to prior work in \Cref{sec:rank_combiner}.
\subsection{The Extractor}
Finally, we are ready to present our explicit quadratic extractor. Given a source $X$, our extractor $\qext(X)$ does the following:
\begin{enumerate}
    \item $W=\{\rcombext(X,i)\}_{i\in [N]}$
    \item $S=\{\qcb(X,W^{i},i)\}_{i\in [N]}$
    \item Output $\nobfext(S)$.
\end{enumerate}
Our analysis splits into two cases based on if the source is pure or not. In the pure case, we first use \Cref{thm:structure_injective} to decompose our source into sources having high derivative entropy. We then use the new linear seeded extractor (\Cref{clry:explicit_rank_combiner_extractor_overview}), our quadratic correlation breaker (\Cref{thm:quad_correlation_breaking}) and the NOBF extractor from \Cref{thm:nobf_extractor} to complete the proof. When the source is not pure, we use the affine analysis and the fact that all the quadratic objects we constructed are easily shown to also work for affine sources. See \Cref{thm:main_theorem_quadratic} for the proof.
\section{Preliminaries}
This section contains useful definitions, results from prior work, and some useful elementary lemmas.
\subsection{Definitions}
We first define the notion of samplable sources.

\begin{definition}[Samplable Source]
    Let $\mathcal{F}$ be a class of boolean functions mapping $s$ bits to $n$ bits. A source $X$ is $\mathcal{F}$-samplable if $X=f(U_s)$ for some $f\in \mathcal{F}$. $f$ is called the generating map and the inputs bits to $f$ are called the generating bits. We use the shorthand $\mathcal{F}$-source.
\end{definition}

We now define \textit{Quadratic sources}, which are the main focus of this paper. 
\begin{definition}[Quadratic sources]
Quadratic sources are the class of $\mathcal{F}$-sources for $\mathcal{F}$ being the class of maps computed by quadratic polynomials over $\F_2$. 
\end{definition}

A $d$-local map is function $f:\F_2^{s}\to\F_2^{n}$ such that $f_i$ is a $d$-junta for each $i\in [n]$. A $d$-local source is a source whose generating map is a $d$-local function. $\nc$ sources are the set of all $d$-local sources for constant $d$.
$\ac$ sources are the set of all sources with the generating map being computed by a low depth polynomial size circuit. 
\begin{definition}
    Let $X=f(U_s)$ be a samplable source. A source $Y$ is a subsource of $X$, denoted by $Y\subset X$, if there exists a subset $S\subseteq \{0,1\}^{s}$ such that $Y=f(U_S)$. Similar to purity \Cref{def:pure_quadratic}, this definition also depends on the generating map $f$.
\end{definition}
Our notion of subsource is non-standard but slightly more convenient (\cite{BKSSW10} defines subsources as obtained by conditioning the distribution $X$ on some event, any subsources from their definition is also a subsource in ours).  Throughout this paper, when we use the term \textit{subsource}, we refer to the above notion.

We now define advice correlation breakers. 
\begin{definition}[Advice correlation breaker]
    \label{defn:correlation_breaker}
A function $\CB:\cube{n}\times \cube{d}\times \cube{a}\to \cube{m}$ is a $(t,k,\epsilon)$-advice correlation breaker if for random variables $X,X^{[t]}, Y,Y^{[t]}$ and advice strings $\alpha,\{\alpha^i\}_{i\in [t]}$ in $\cube{a}$ satisfying
\begin{enumerate}
    \item $\hinf(X)\geq k$, 
    \item $Y=U_d$,
    \item $X,X^{[t]}$ is independent of $Y,Y^{[t]}$,
    \item $\alpha\neq \alpha^i$ for all $i\in [t]$, 
\end{enumerate}
it holds that 
\begin{equation}
    \label{eqn:correlation_breaker_guarantee}
    (\CB(X,Y,\alpha)\approx_{\epsilon} U_m)\mid \{\CB(X^{i},Y^{i},\alpha^{i})\}_{i\in [t]}
\end{equation}where $U_m$ is uniform random string independent of $X,X^{[t]},Y,Y^{[t]}$.
\end{definition}
\begin{definition}[$t$-wise independence]
    A distribution $X$ on $\F_2^{n}$ is $t$-wise independent if $X_S=U_{|S|}$ for all sets $S$ satisfying $|S|\leq t$.
\end{definition}

\subsection{Results from Prior Work}
We need a linear seeded extractor due to \cite{chattopadhyay2021affine}, which has good seed length when the output is small.
\begin{theorem}[\cite{chattopadhyay2021affine}]
    \label{thm:cglext}
    For any $n,\epsilon,m$, there exists an explicit $(m+2\log(1/\epsilon),\epsilon)$ strong seeded extractor $\cglext:\bcube{n}\times \bcube{d}\to \bcube{m}$ with $d=2m+8\log(n/\epsilon)+O(1)$. In particular for $m=O(\log(n/\epsilon))$, the seedlength is $O(\log (n/\epsilon))$.
\end{theorem}


We also need a linear seeded extractor from \cite{li2016improved}, which has good seed length when the error is large.

\begin{theorem}[\cite{li2016improved}]
    \label{thm:li_lext}
    There is a constant $c>1$ such that for all $n,k$ such that $c\log^8 n\leq k\leq n$, and $\epsilon\geq n^{-2}$, there exists an explicit $(k,\epsilon)$ strong linear seeded extractor $\lext:\bcube{n}\times \bcube{d}\to \bcube{m}$ with $d=O(\log n)$ and $m=\sqrt{k}$. 
\end{theorem}


The following lemmas measure closeness to uniform and $t$-wise independent distributions respectively.
\begin{fact}[\cite{vazirani1986randomness}]
    \label{fact:vazirani_xor}
    Let $X$ be a random variable over $\bcube{n}$. Suppose $|\E[\chi_S(X)]|\leq \epsilon$ for all non-empty $S\subseteq [n]$. Then, $X$ is $\epsilon\cdot 2^{n/2}$-close to uniform. 
\end{fact}
\begin{fact}[\cite{agm03}]
    \label{fact:agm}
    Let $X$ be a random variable over $\bcube{n}$. Suppose $|\E[\chi_S(X)]|\leq \epsilon$ for all non-empty $S\subseteq [n]$ satisfying $|S|\leq t$. Then, $X$ is $2n^{t}\epsilon$-close to $t$-wise independent. 
\end{fact}

We need the following affine extractor due to Li \cite{li2016improved}.
\begin{theorem}[{\cite[Theorem 1.7]{li2023two}}]
    \label{thm:affine_extractor}
    There exists a constant $C>0$ and function $\affinext:\cube{n}\to\cube{1}$ such that for all affine sources $X$ with $\hinf(X)\geq C\log n$, it holds that 
    $\dtv(\affinext(X),U_1)\leq 0.1$.
\end{theorem}


We also need the following claim, which shows that a strong linear seeded extractor with overall error \(\epsilon\) automatically achieves closeness error \(0\) and strength error \(2\epsilon\) on affine sources.


\begin{lemma}[\cite{rao2009extractors}]
    \label{lem:rao_affine_lext}
    Let $\lext:\bcube{n}\times \bcube{d}\to \bcube{m}$ be a $(k,\epsilon)$ strong linear seeded extractor with $\epsilon\in [0,1]$. Then, for any affine source $X$ with entropy $k$, it holds that 
    \[
    \Pr_{s\in \bcube{d}}[\dtv(\lext(X,s),U_m)=0]\geq 1-2\epsilon.    
    \]
    That is, $\lext$ is a $(k,\epsilon,0)$-strong linear seeded extractor for affine sources.
\end{lemma}

The following is a corollary of \Cref{thm:cohen-tal}. 
\begin{corollary}
    \label{clry:cohen_tal}
    Given any quadratic polynomials $q_1,\ldots,q_t$ from $\bcube{n}\to\{0,1\}$, there exists a universal constant $C$ and a function $\ctpartition:\bcube{n}\to \bcube{n-d}$ for $d=C(n/t)-t$ such that for all $c\in \bcube{n-d}$, it holds that $q_1,\ldots,q_t$ is constant on the set $\ctpartition^{-1}(c)$ and this set is an affine subspace.
\end{corollary}
\begin{proof}
    Index the subspaces in the partition given by \Cref{thm:cohen-tal} in arbitrarily. Note that there are exactly $2^{n-d}$ such (affine) subspaces. Now, the function $\ctpartition$ just outputs the index of the subspace that a given input belongs to.
\end{proof}

We also use the following strong $t$-advice correlation breaker from \cite{li2023two}.

\begin{theorem}[Explicit advice correlation breaker \cite{li2023two}]
    \label{thm:li_correlation_breaker}
    For any positive integer $t$, there exists an explicit strong $(t,k,\epsilon)$-advice correlation breaker $\CB:\cube{n}\times \cube{d}\times \cube{a}\to \cube{m}$ for $k=O(tm+t\log(n/\epsilon)\cdot\frac{\log a}{\log \log a}+t^2\log(n/\epsilon))$, with $d=O(tm+t\log(n/\epsilon)\cdot \frac{\log a}{\log \log a}+t\log^3(t)\log(n/\epsilon))$.
\end{theorem}

The following input reduction lemma is due to \cite{polysourceextcgg}.
\begin{lemma}[Input reduction: entropy version]
    \label{lem:input_reduction_entropy}
Let $X=f(U_s)$ be a quadratic source such that $\hinf(X)\geq k$. For any $\epsilon>0$, there exists a full rank linear map $L:\F_2^{s}\to\F_2^{s-\ell}$ with $\ell\leq k+O(\log(1/\epsilon))$ such that 
\[
\Pr_{b\sim L(U_s)}[\hinf^{\epsilon}(f(U_s)\mid L(U_s)=b)\geq k-1]\geq 1-\epsilon.
\]
In particular, $X$ is $\epsilon$-close to a convex combination of quadratic sub-sources $X'$ with $\hinf^{\epsilon}(X')\geq k-1$ and at most $k+O(\log(1/\epsilon))$ generating bits.
\end{lemma}
\begin{proof}
We borrow notation from the proof of Lemma~2.4 in \cite{polysourceextcgg}. Let $\ell=k+O(\log(1/\epsilon))$. Their proof shows that there exists a full rank linear map $L:\F_2^{s}\to \F_2^{s-\ell}$ and a function $S:\bcube{n}\to \bcube{k+1}$ such that (1) $\hinf(S(X))\geq k-1$ and (2) with probability at least $1-\epsilon$ over $b\sim L(U_s)$, the source $\tilde{X}=(f(U_s)\mid L(U_s)=b)$ satisfies 
\[
    \dtv(S(\tilde{X}),S(X))\leq \epsilon.
\]

From the fact that $L$ is full rank and has codimension $\ell$, we have that $\tilde{X}$ is a quadratic source with at most $\ell$ generating bits. 

Now, all that remains to prove the lemma is to show that $\tilde{X}$ is $\epsilon$ close to a source $X'$ with $\hinf(X')\geq k-1$. We now construct such a $X'$. Define $\tilde{X}_s$ as the distribution as the conditional distribution $(\tilde{X}\mid S(\tilde{X})=s)$ (for $s\notin \supp(S(\tilde{X}))$, define $\tilde{X}_s$ as an arbitrary distribution over $S^{-1}(s)$). Define $X'$ as follows: (1) sample $s\sim S(X)$, and (2) sample $X'\sim \tilde{X}_s$. From data processing inequality, we have that 
\[
\dtv(\tilde{X},X')\leq \dtv(S(\tilde{X}),S(X))\leq \epsilon.
\]

Now, for any string $x$, we have $\Pr_{X'}[X'=x]\leq \Pr_{X'}[S(X')=S(x)]=\Pr_{X}[S(X)=S(x)]\leq 2^{-k+1}$. This proves that $\hinf(X')\geq k-1$.
\end{proof}

\subsection{Additional Lemmas}
We recall that a source $X$ has $\epsilon$-smooth min entropy $k$, denoted by $\hinf^{\epsilon}(X)\geq k$ if there exists some source $\tilde{X}$ with $\hinf(\tilde{X})\geq k$ such that $X\approx_{\epsilon} \tilde{X}$.
We will need the following smooth min-entropy chain rule. We include the proof here for completeness. 
\begin{lemma}
    \label{lem:chain_rule_smooth}
    Let $(X,Y)$ be random variables such that $\hinf^{\epsilon}(X)\geq k$ and $|\supp(Y)|\leq M$. Then, it holds with probability at least $1-(\epsilon+2\delta)$ over $y\sim Y$ that 
    \[
    \hinf^{2\epsilon/\delta}(X\mid Y=y)\geq k-\log M-\log(1/\delta).    
    \]
\end{lemma}
\begin{proof}
    Let $Y_x$ be the conditional distribution of $Y$ conditioned on $X=x$. Let $\tilde{X}$ be the distribution such that $\hinf(\tilde{X})\geq k$ and $\dtv(X,\tilde{X})\leq \epsilon$. Let $\tilde{Y}$ be the random variable obtained by first sampling $x\sim \tilde{X}$ and then sampling $\tilde{Y}\sim Y_x$ (define $Y_x$ arbitrarily over $\supp(Y)$ for $x\notin \supp(X)$).

    From min-entropy chain rule, we have that with probability at least $1-\delta$ over $y\sim\tilde{Y}$, it holds that $\hinf(\tilde{X}\mid \tilde{Y}=y)\geq k-\log M-\log(1/\delta)$,
    By data processing inequality, we also have that $\dtv((X,Y),(\tilde{X},\tilde{Y}))\leq \epsilon$. This implies that
    \begin{align*}
        \E_{y\sim Y}\dtv((X\mid Y=y), (\tilde{X}\mid \tilde{Y}=y))\leq \dtv((X,Y),(\tilde{X},\tilde{Y}))+\dtv(Y,\tilde{Y})\leq 2\epsilon.
    \end{align*}
    From Markov's inequality, we have that with probability at least  $1-\delta$ over $y \sim Y$, we have $\dtv((X\mid Y=y), (\tilde{X}\mid \tilde{Y}=y))\leq 2\epsilon/\delta$. Using the fact that $\dtv(Y,\tilde{Y})\leq \epsilon$ and $\Pr_{y\sim \tilde{Y}}[\hinf(\tilde{X}\mid \tilde{Y}=y)< k-\log M-\log(1/\delta)]\leq \delta$, we have that with probability $1-(\epsilon+2\delta)$ over $y\sim Y$, 
    \[
    \hinf^{2\epsilon/\delta}(X\mid Y=y)\geq k-\log M-\log(1/\delta).    
        \]
\end{proof}

\begin{lemma}
    \label{lemma:smooth_entropy_slice}
    Let $X$ be a distribution on $\bcube{n}$ such that $\hinf^{\epsilon}(X)\geq n-\Delta$. Then, $\hinf^{\epsilon}(X_{[d]})\geq d-\Delta$.
\end{lemma}
\begin{proof}
By definition of smooth min-entropy, there is some \(\tilde{X}\approx_\epsilon X\) with \(H_\infty(\tilde{X})\geq n-\Delta\). By \cite[Claim 4.11]{polysourceextcgg}, $\hinf(\tilde{X}_{[d]})\geq d-\Delta$. By the data-processing inequality, \(X_{[d]}\approx_\epsilon \tilde{X}_{[d]}\), which completes the proof.
\end{proof}

The following lemma shows that for affine sources, smooth min-entropy is close to the true min-entropy.
\begin{lemma}
    \label{lem:smooth_entropy_to_entropy}
    Let $X$ be a flat source satisfying $\hinf^{1/2}(X)\geq k$. Then, it holds that $\hinf(X)\geq k-1$
\end{lemma}
\begin{proof}
    Let $S=\supp(X)$ and $f$ be the indicator function on this set. Let $\tilde{X}$ be the distribution satisfying $\hinf(\tilde{X})\geq k$ and $\tilde{X}\approx_{1/2}X$ guaranteed by the premise of the lemma. We have that $\E[f(\tilde{X})]\geq 1/2$ (from the TV distance guarantee) and $\E[f(\tilde{X})]\leq |S|/2^{k}$ (from the min-entropy guarantee). 
    Thus, we have that $|S|\geq 2^{k-1}$ which implies $\hinf(X)\geq k-1$.
\end{proof}

The next lemma is useful in the proof of our structure theorem \Cref{thm:structure_subseed}.
\begin{lemma}[Weak-injectivity preservation]
\label{lem:weak_injectivity_preservation}
    Let $X=f(U_{k+\ell})$ be a random variable satisfying $\hinf^{\epsilon}(X)\geq k$ for $f:\F_2^{k+\ell}\to\F_2^{n}$. Let $\mathcal{S}$ be a distribution over subsets of $\F_2^{k+\ell}$ such that (1) $\supp(\cal{S})$ forms a partition of $\F_2^{k+\ell}$ and (2) for every set $s\in \supp(\mathcal{S})$, it holds that $\Pr_{S\sim \mathcal{S}}[S=s]=|s|/2^{k+\ell}$. Then, with probability $1-\delta$ over $S\sim \mathcal{S}$, it holds that
    \[
    \hinf^{\epsilon/\delta}(f(U_S))\geq \log(|S|)-\ell.
    \]
\end{lemma}
\begin{proof}
    Let $A=\{x\mid |f^{-1}(x)|\geq 2^{\ell}, x\in \F_2^{n}\}$. By the smooth min-entropy guarantee, we have that 
    \[
    |f^{-1}(A)|/2^{k+\ell}=\Pr[X\in A]\leq |A|/2^{k}+\epsilon.
    \]
    This implies that $|f^{-1}(A)|-|A|2^{\ell}\leq \epsilon2^{k+\ell}$. For each $x\in A$, choose exactly $2^{\ell}$ entries in $f^{-1}(x)$ arbitrarily. Let $B$ be the elements in $f^{-1}(A)$ that are not chosen. We have that $|B|\leq \epsilon2^{k+\ell}$. Define a function $g$ that agrees with $f$ on $\F_2^{k+\ell}\setminus B$ and assigns the points in $B$ to $\F_2^{n}$ in such a way that $|g^{-1}(x)|\leq 2^\ell$ for all $x\in \F_2^{n}$.

    Now, it is easy to see that $\hinf(g(U_S))\geq \log(|S|)-\ell$ for all sets $S$. Also, using the fact that $\Pr[U_{k+\ell}\in B]=\E_{S\sim \mathcal{S}}[\frac{|S\cap B|}{|S|}]\leq \epsilon$, we have that with probability at most $1-\delta$ over $S\sim \mathcal{S}$, it holds that $\frac{|S\cap B|}{|S|}\leq \epsilon/\delta$. Now, since $f$ and $g$ agree outside $B$, we have $g(U_S)\approx_{\epsilon/\delta}f(U_S)$. Combining this with the fact that $\hinf(g(U_S))\geq \log(|S|)-\ell$ completes the proof. 
\end{proof}
We use the following lemma that states that pure-quadratics remain pure after restricting to subsources formed by affine subspaces of the generating bits.
\begin{lemma}
\label{lem:purity_preserving}
    Let $X=f(U_s)$ be a $q$-pure quadratic source. Then $X'=(f\circ L)(U_r)$ is also $q$-pure for any injective affine map $L$.
\end{lemma}
\begin{proof}
Let $V\subseteq \F_2^{r}$ be an affine subspace such that $(f\circ L)|_{V}$ is affine with $\dim((f\circ L)(V))\geq q$. This implies that the purity of $f$ is violated on the subspace $L(V)$.
\end{proof}
\section{Structure Theorem}
\label{sec:structure_theorem}
We now  show a structure theorem for quadratic sources. 
\subsection{Injective Sources}
We present the proof of \Cref{thm:structure_injective} here for completeness.
\begin{proof}
    We will decompose the source by iteratively restricting into affine subspaces of the generating bits.     
    We will execute the decomposition in $t$ stages. 
    We maintain the following invariant: at the end of stage $j$, we will ensure that our original source will be exactly decomposed into a convex combination of subsources such that 
    \begin{enumerate}
        \item each subsource $W$ has the form
    $W=h(Y,Z)=g(Y)+\sum_{i=1}^{j'}b_i Z_i +$ where $Y=U_{k'}$, $Z=U_{j'}$  (for $k'\geq k-rj'$ and $j'\leq j$) are uniform and independent,
    \item  $j'\neq j$ implies that $\hinf(\partial_a g(Y))\geq r$ for all $a\neq 0$, and 
    \item $h$ is injective. 
    \end{enumerate}

    We now describe our iterative decomposition procedure. Note that the invariant trivially holds at the start ($j=0$). We now describe stage $j+1$, assuming that the invariant holds after stage $j$. 

    Consider any subsource $W=g(Y)+\sum_{i=1}^{j'}b_i Z_i$ obtained after stage $i$. First, we leave the subsource unchanged if it holds that $\hinf(\partial_a g(Y))\geq r$ for all $a\neq 0$. If the previous property does not hold, then it holds from item (2) of the invariant that $j'=j$. Let $a\neq 0$ be a direction along which $\hinf(\partial_a g(Y))< r$. We will further decompose $W$ by conditioning on the value of $\partial_a g(Y)$. Consider the subsource obtained from $W$ by conditioning on the event  $\{\partial_a g(Y)=b\}$. 
    
    Let $A=c+B\subseteq \bcube{k'}$ be the (affine) subspace such that $\partial_a g(y)=b$ for $y\in A$ where $B$ is a subspace and $c$ is a fixed vector. We know that $\dim (A)\geq |Y|-(r-1)= k-rj-(r-1)$ as $\dim(\partial_a g(\cdot))<r$. Observe that $a+A=A$ which implies $a\in B$. Starting with $a$, construct an arbitrary basis $\{a, e_1,\ldots, e_{\dim(B)-1}\}$ of $B$. Let $C=c+\mathrm{span}(\{e_i\}_{i\in [\dim(B)]-1})$. Let $Y'$ be a uniform string from $A$. We have that $Y'=Za + U_{C}$ where $Z$ is a uniform bit and $U_{C}$ is independent of $Z$ and uniform over $C$. 
    
    The subsource $W'$ obtained from $W$ by conditioning on the derivative is exactly $W'=g(Y')+\sum_{i=1}^{j'}b_i Z_i $. Note that $g(y'+a)=g(y')+b$ for all $y'\in A$. Hence, we have that $W'=g(U_C)+\sum_{i=1}^{j} b_i Z_i+ Zb$. We have that $\dim(C)=\dim (A)-1\geq k-r(j+1)$. Thus $E=L(U_{k''})$ for some full rank affine map $L$ and $k''\geq k-r(j+1)$. 
    
    Defining $Z_{j+1}=Z$, $b_{j+1}=b$ and $g'=g\circ L$ shows that $W'$ has the form $g'(U_{k''})+\sum_{i=1}^{j+1}b_iZ_i$. Note that the injectivity of the map is preserved as we started out with an injective map and $Y'$ is uniform over a subset of inputs. Thus, $W'$ satisfies the invariant after stage $j+1$. 
    
     We will execute the procedure for $j=t$ stages. Consider any subsource $W$ in the decomposition obtained after $t$ stages. Recall $W=h(Y,Z)=g(Y)+\sum_{i=1}^{j'}Z_ib_i$. There are two cases. In the first case, we have that $j'=t$. In this case, we can further fix the random variable $Y$ to obtain an affine source $W'=\sum_{i=1}^{t}Z_ib_i + d$ where $d$ is a constant shift. Since $h$ was injective, this subsource must have entropy $t$. In the second case, we have that $\hinf(\partial_a g(Y))\geq r$ for all $a\neq 0$. In this case, we further fix all the variables $Z$ to obtain a source of the second type in the theorem statement. 
\end{proof}
\subsection{General Quadratic Sources}
We now extend \Cref{thm:structure_injective} to general quadratic sources with a smooth min-entropy guarantee with the number of generating bits slightly larger than the entropy. Our decomposition procedure only acts on a subset of the generating bits that we choose. 

\begin{theorem}
    \label{thm:structure_subseed}
        Let $X=f(Y,Z)$ be a quadratic source with $\hinf^{\epsilon^3}(X)\geq k$ and generating map $f:\bcube{k+\ell}\to \bcube{n}$ where $Y=U_s$ and $Z=U_{k+\ell-s}$. Let $r,t$ be positive integers such that $rt\leq s$. Then $X$ is $\epsilon$ close to a convex combination of subsources of the form $f(U_V)$ for affine subspace $V\subseteq \F_2^{k+\ell}$ such that the subsources are one of the following two types: 
        \begin{enumerate}
            \item Affine sources $W$ with min-entropy at least $t-\ell-1$ ($f|_{V}$ is affine) or
            \item Quadratic sources $W(Y',Z)=g_1(Y')+g_2(Y',Z)+g_3(Z)$ such that (1) $Y'=U_{s'}$ for $s'\geq s-rt$, (2) $g_2$ is bilinear, (3) $\hinf(\partial_a g_1(Y'))\geq r$ for all $a\neq 0$ and (4) $\hinf^{\epsilon^2}(W(Y',Z))\geq s'+|Z|-\ell$ . 
        \end{enumerate}
        Furthermore, $Z$ is the same seed random variable as that in the source $X$ and is independent of $Y'$. 
    \end{theorem}
    \begin{proof}
       
        As in the proof of \Cref{thm:structure_injective}, we will iteratively decompose the source into sub-sources. 
        A crucial difference is that we will only condition on functions of the input seed $Y$, leaving $Z$ unchanged. We maintain the following invariant. 
    
        After $j$ stages, we ensure that all the component subsources $W$ in the convex combination satisfy the following: 
        \begin{enumerate}
            \item $W=h(Y'=(A,B),Z)=g_1(B)+\sum_{i=1}^{j'}A_i\cdot (b_i(Z))+g_2(B,Z)+g_3(Z)$ where $b_i(Z)$ are affine functions of $Z$ for all $i\in [j']$ and $g_2$ is bilinear (here $j'\leq j$).
            \item $A=U_{j'}$ and $B=U_{s'}$ for $s'\geq s-rj'$. $A$ and $B$ are independent of each other and $Z$.
            \item Furthermore $j'\neq j$ implies that  $\hinf(\partial_a g_1(.))\geq r$ for all $a\neq 0$.
        \end{enumerate}
    
        Observe that the invariant trivially holds for $j=0$ (with $|A|=0$). We will now describe stage $j+1$. Assume the invariant holds upto stage $j$. Take a component $W$ obtained after $j$ stages (satisfying the above invariant). We check if all the non-zero derivative entropies of $g_1$ are greater than equal to $r$. If so, we make no changes to $W$ (this component is frozen in future iterations as well). Alternatively, if there is a direction $a$ with low derivative entropy, we choose the lexicographically smallest such direction and fix the value of this derivative. 
        
        Suppose we condition on the event $\{\partial_a g_1(B)=b\}$. Observe that $\partial_a g_2(B,Z)=g_2(a,Z)$ which is a function of $Z$. Thus, after fixing the derivative, we can retrieve one "affine" direction (if one treats $Z$ as a constant). That is, using the same argument as the proof of \Cref{thm:structure_injective}, the new subsource can be expressed as $W'=h(Y'=(A,D),Z,B')=g_1(B')+\sum_{i=1}^{j'}A_i\cdot (b_i(Z))+D (g_2(a,Z)+b)+g_2(B',Z)+g_3(Z)$ where $D$ is a fresh random bit, and $B'+Da=B\mid \{\partial_a g_1(B)=b\}$. Rewrite $B'$ as $L(U_{s''})$ for a full rank affine function with $s''\geq s'-r\geq s-r(j'+1)$. Now, we observe that $W'$ satisfies the invariant. 

        We run the decomposition for $j=t$ stages.  Let the subsource obtained after the $t$ stages be $W=h(Y'=(A,B),Z)=g_1(B)+\sum_{i=1}^{j'}A_i\cdot(b_i(Z))+g_2(B,Z)+g_3(Z)$. 
        
        There are again two cases. In the first case, we have $j'=t$. In this case, we fix $B,Z$ to obtain an affine source. In the second case, we have $\hinf(\partial_a g_1(Y'))\geq r$ for all $a\neq 0$. In this case, we further fix all the variables $A$ to obtain a source of the form described in Item~2 of the theorem statement. 

        We now argue that the entropy guarantees stated in both the cases are satisfied (recall that this was immediate in the injective case). To do this, we use \Cref{lem:weak_injectivity_preservation}. First, observe that all the subsources obtained by this decomposition are of the form $f(U_V)$ for some affine subspace $V\subseteq \F_2^{k+\ell}$. Also, the (affine) subspaces corresponding to different subsources form a partition on $\F_2^{k+\ell}$. In the branches of the decomposition leading to affine subspaces (Item~1), we have $\dim(V)=t$ and in the other case, we have $\dim(V)=s'+|Z|$. Let $\mathcal{V}$ be a random variables that chooses these affine subspaces according to the above decomposition. From \Cref{lem:weak_injectivity_preservation}, we have that 
        with probability $1-\epsilon$ over $V\sim \mathcal{V}$, it holds that (1) $\hinf^{\epsilon^2}(f(U_V))\geq t-\ell$ if $V$ came from decompositions in the first case, or (2) $\hinf^{\epsilon^2}(f(U_V))\geq s'+|Z|-\ell$ in the second case. Applying \Cref{lem:smooth_entropy_to_entropy} to the first case implies the entropy guarantees in the theorem statement. 
    \end{proof}

We now state an immediate corollary of the above theorem that gives a sufficient condition for decomposing quadratic sources into sub-sources with high derivative entropy. 

\begin{corollary}[Pure Quadratics can be decomposed into high derivative entropy sources]
\label{clry:pure_quadratic_structure}
      Let $X=f(Y,Z)$ be a quadratic source with $\hinf^{\epsilon^3}(X)\geq k$ and generating map $f:\bcube{k+\ell}\to \bcube{n}$ where $Y=U_s$, $Z=U_{k+\ell-s}$. Let $r,t$ be such that $rt\leq s/2$ and suppose $X$ is $(t-\ell-1)$-pure. Then, $X$ is $\epsilon$ close to a convex combination of subsources of the form 
      \[
      X'=g(Y',Z)=g_1(Y')+g_2(Y',Z)+g_3(Z)
      \]
      such that (1) $\hinf(\partial_a g_1(Y'))\geq r$ for all $a\neq 0$, (2) $\hinf^{\epsilon^2}(X')\geq s'+|Z|-\ell$, (3) $Y'=U_{s'}$ with $s'\geq s/2$, and (4) $g_2$ is bilinear. Again, the random variable $Z$ is left unchanged by the decomposition and $X'$ is obtained by restricting $f$ to an affine subspace of the generating bits.
\end{corollary}
\begin{proof}
    The proof is immediate from \Cref{thm:structure_subseed} as Item~1 in \Cref{thm:structure_subseed} is not possible anymore.
\end{proof}

 \section{Exploiting High Derivative Entropy}
The below argument shows that a linear seeded extractor extracts from a high derivative entropy quadratic source with a quadratically correlated seed.

\begin{lemma}
    \label{thm:cauchy_schwarz_general}
        Let $X=f(Y,Z)$ be a source with $f$ quadratic, $Y=U_{s}$, $Z\in\F_2^{t}$ independent of $Y$ and such that $f(Y,Z)=f_1(Y)+f_2(Y,Z)+f_3(Z)$ where $f_2$ is bilinear and $\hinf(\partial_a f_1(Y))\geq k$ for all $a\neq 0$. Let $\lext:\bcube{n}\times \bcube{d}\to \bcube{m}$ be a $(k,\epsilon)$ strong linear seeded extractor. Then, for any random variable $W$ (correlated with $Z$, independent of $Y$) satisfying $W\approx_{\delta}U_d$, it holds with probability $1-2^{m}\cdot (2^{-s}+2\epsilon+\delta)^{1/2}$ over $(z,w)\sim (Z,W)$ that 
        \[
        \dtv(\lext(f(Y,z),w),U_{m})\leq 2^{m/2}(2^{-s}+2\epsilon+\delta)^{1/4}.
        \]
        
       Furthermore, the theorem also holds when $W\approx_{\delta}U_d$ is replaced by $\hinf^{\delta}(W)\geq d-\Delta$ with $2\epsilon+\delta$ replaced by $2^{\Delta+1}\epsilon+\delta$ in both the probability and TV distance bounds.
    \end{lemma}
    \begin{proof}
         We will first argue that for any $S\subseteq [m]$, it holds high probability over $(z,w)\sim (Z,W)$ that $|\E[\chi_S(\lext(f(Y,z),w))]|\leq \epsilon$. The theorem then follows by a union bound over all subsets $S$ and \Cref{fact:vazirani_xor}.
    
        Let $b(w,z)=|\E_{Y}[\chi_S(\lext(f(Y,z),w))]|$. We have that 
        \begin{align*}
            \E_{W,Z}[(b(W,Z))^2]&=\E_{W,Z}(\E_{Y}[\chi_S(\lext(f(Y,Z),W))])^2\\
            &=\E_{W,Z}\E_{Y,A}[\chi_S(\lext(f(Y,Z)+f(Y+A,Z),W))]\\
            &=\E_{W,Z}\E_{Y,A}[\chi_S(\lext(\partial_A f_1(Y)+f_2(A,Z),W)).
        \end{align*}
        In the second inequality $A$ is uniform and independent of $Y$.
        Now for every $a\neq 0$, we have $\hinf(\partial_a f_1(Y))\geq k$ and $\partial_a f_1(Y)$ is an affine source. From \Cref{lem:rao_affine_lext}, we have that with probability at least $1-2\epsilon-\delta$ over uniform string $w\sim W$, it holds that $\dtv(\lext(\partial_a f_1(Y),w),U_{m})=0$. 
    Thus, we have 
    \begin{align*}
          \E_{W,Z}[(b(W,Z))^2]&=\E_{W,Z}\E_{Y,A}[\chi_S(\lext(\partial_A f_1(Y)+f_2(A,Z),W))]\\
          &\leq \Pr[A=0]+\max_{a\neq 0}\E_{Y,W,Z}[\chi_S(\lext(\partial_a f_1(Y)+f_2(a,Z),W))]\\
          &\leq 2^{-s}+\max_{a\neq 0}\E_{W,Z}|\E_{Y}[\chi_S(\lext(\partial_a f_1(Y),W))]|\\
          &\leq 2^{-s}+2\epsilon+\delta.
        \end{align*}
        Now, by Markov's inequality, we have that $b(w,z)\leq (2^{-s}+2\epsilon+\delta)^{1/4}$ with probability $1-(2^{-s}+2\epsilon+\delta)^{1/2}$. 
       The theorem statement now follows from a union bound and \Cref{fact:vazirani_xor}. For the guarantee assuming high smooth min-entropy, the probability of sampling a bad seed is at most $2^{\Delta+1}\epsilon+\delta$ and the rest of the proof is the same.

    \end{proof}

\section{Quadratic Correlation Breaking}
\label{sec:quadratic_correlation_breaking}
In this section we construct a new type of correlation breaker that can handle quadratic sources and seeds. This can be viewed as a generalization of an affine correlation breaker. The construction we use is from \cite{chattopadhyay2022extractors}, where they show how to upgrade a black box advice correlation breaker into an affine correlation breaker. We will show that this construction is also a \textit{quadratic correlation breaker} when the source is a pure-quadratic. 

We first define the notion of a quadratic correlation breaker.
\begin{definition}[Quadratic Correlation Breaker]
    \label{defn:quadratic_correlation_breaker}
A function $\qcb:\F_2^{n}\times \F_2^{d}\times \cube{a}\to \cube{m}$ is a $(t,k,q,\delta)$-quadratic correlation breaker if there exists an absolute constant $c>0$ such that for random variables $X,W,W^{[t]}$ and advice strings $\alpha,\{\alpha^i\}_{i\in [t]}$ in $\cube{a}$ satisfying
\begin{enumerate}
    \item $X=f(V)$ is a $q$-pure quadratic source with $\hinf(X)\geq k$ and $V=U_s$ where $s\leq k+c\log(1/\delta)$,
    \item $W=p(V)$, $W^{i}=p^{i}(V)$ for all $i\in [t]$ with $p,p^1,\ldots,p^t:\F_2^{s}\to\F_2^{d}$ are quadratic maps, 
    \item $W=U_d$,
    \item $\alpha\neq \alpha_i$ for all $i\in [t]$, 
\end{enumerate}
it holds that 
\begin{equation}
    \label{eqn:qcb_guarantee_main}
    (\qcb(X,W,\alpha)\approx_{\delta} U_m)\mid \{\qcb(X,W^{i},\alpha^{i})\}_{i\in [t]}
\end{equation}where $U_m$ is uniform random string independent of $X,W^{[t]}$.
\end{definition}

    We now describe the construction of our quadratic correlation breaker (as mentioned before, the construction already appears in \cite{chattopadhyay2022extractors}). 
    \begin{algorithm}
        \caption{Quadratic Correlation Breaker ($\qcb$)}
        \label{alg:quadratic_correlation_breaker}
       \begin{algorithmic}
        \STATE \textbf{Input: } $x\in \{0,1\}^{n}$--- an $n$-bit string, $w\in \cube{d}$--- the seed, $\alpha\in \cube{a}$--- the advice string.    
        \STATE \textbf{Output:} $s\in \{0,1\}^{m}$--- output string.
        \STATE \hrulefill
       \STATE  \textbf{Sub-Routines and Parameters:}
       \STATE Let $ d_1=O(t(m+t\log a\log(n/\delta))), d_0=O(d_1+\log(1/\delta)), d=O(t^5(m+\log a\log(n/\delta))^2)$.
       \STATE Let $\lext:\cube{n}\times \cube{d_0}\to \cube{d_1}$ be a $(k,2^{-4d_1} \delta^{32})$ strong linear seeded  from \Cref{thm:cglext}.
       \STATE Let $\CB:\cube{d}\times \cube{d_1}\times \cube{a}\to \cube{m}$ be a $(t,d/2,\delta/2)$-advice correlation breaker from \Cref{thm:li_correlation_breaker}.
       \STATE \hrulefill
        \begin{enumerate}
            \item Let $s_1=w_{[d_0]}$.
            \item Let $r=\lext(x,s_1)$.
            \item Let $s=\CB(w,r,\alpha)$
        \end{enumerate}
    \end{algorithmic}    
    \end{algorithm}

We now state the main theorem of this section.
    \begin{theorem}
    \label{thm:quad_correlation_breaking}
      For any $\delta\in (0,1)$ and positive integers $t,a,m$, there exists a $(t,k,q,O(\delta))$-quadratic correlation breaker $\qcb:\F_2^{n}\times \F_2^{d}\times \F_2^{a}\to\F_2^{m}$ (from \Cref{alg:quadratic_correlation_breaker}) whenever
      \begin{itemize}
          \item $d\geq Ct^{5}(m+\log a\log(n/\delta))^2$,
          \item $k\geq Ct^2d^4$,
          \item $q\leq c\sqrt{k/(td^2)}$,
      \end{itemize}
      for absolute constant $c,C>0$. Moreover, the guarantee in \Cref{eqn:correlation_breaker_guarantee} also holds when the entropy assumption in Item~1 and the uniformity assumption in Item~3 of \Cref{defn:quadratic_correlation_breaker} are relaxed to $\hinf^{\delta^4}(X)\geq k$ and $W\approx_{\delta^2}U_d$ respectively.
    \end{theorem}

   Before proving \Cref{thm:quad_correlation_breaking}, we establish an intermediate lemma that proves that the correlation breaker works when the source is decomposed in a specific form. 
    \begin{lemma}
    \label{thm:intermediate_qcb}
 Let $\delta$ be the error parameter in \Cref{alg:quadratic_correlation_breaker}. Let $X=f(Y,Z)$ be a source with quadratic $f$, $Y=U_{s}$, $Z$ independent of $Y$ and such that $f(Y,Z)=f_1(Y)+f_2(Y,Z)+f_3(Z)$ where $f_2$ is bilinear and $\hinf(\partial_a f_1(Y))\geq k$ for all $a\neq 0$. Also let $W,W^{[t]}$ be functions of $Z$ satisfying $W=U_d$. Also, let $\alpha,\alpha^{[t]}$ be such that $\alpha\neq \alpha_i$ for all $i\in [t]$. Suppose $s,k\geq C(d_1+\log(1/\delta))$ for a large constant $C$. Then, it holds that
\[
    \label{eqn:quad_correlation_breaker_guarantee_intermediate}
        (\qcb(X,W,\alpha)\approx_{O(\delta)} U_m)\mid \{\qcb(X,W^{i},\alpha^{i})\}_{i\in [t]}.
\]   \end{lemma}

We need a stronger version of the input reduction lemma to prove the above lemma. 
\begin{lemma}[Input Reduction: Partial Input]
    \label{lem:partial_input_reduction}
    Let $f:\bcube{s+t}\to \bcube{n}$ be a quadatic map. Furthermore, suppose it holds that $\dtv(f(X,Y),f(X,y))\leq \delta$ with probability at-least $1-\delta$ over $y\sim Y$ where $X=U_s$ and $Y$ is independent of $X$. For any $0<\epsilon<1/4$, there exists a full rank linear map $L:\bcube{s}\to \bcube{s-\ell}$ with $\ell=n+O(\log(1/\epsilon))$ such that 
    \[
        (f(X,Y),Y,L(X))\approx_{\epsilon+2\delta} (f(X,Y),Y',U_{s-\ell})  
    \]where $U, X,Y,Y'$ are independent and $Y,Y'$ are identically distributed. 
\end{lemma}
\begin{proof}
    We choose $L$ to be a random linear map. 
    By triangle inequality, we have that for any $L$, it holds that
    \begin{align*}
        \dtv((f(X,Y),Y,L(X)),(f(X,Y),Y',U_{s-\ell}))&\leq \dtv((f(X,Y),Y,L(X)),(f(X,Y),Y,U_{s-\ell}))\\
        &+ \dtv((f(X,Y),Y,U_{s-\ell}),(f(X,Y),Y',U_{s-\ell}))
    \end{align*}
    We bound the first term using \Cref{lem:input_reduction_short}. That is for every fixed $y\sim Y$, we can apply \Cref{lem:input_reduction_short}. By averaging over $y\sim Y$ and $L$ and then using Markov's inequality, we have that $\dtv((f(X,Y),Y,L(X)),(f(X,Y),Y,U_{s-\ell}))\leq \epsilon/2$ with probability at least $1-\epsilon$ over $L$. The second term is at most $2\delta$ from the premise of the lemma that $\dtv(f(X,Y),f(X,y))\leq \delta$ with probability $1-\delta$ over $y\sim Y'$. 
\end{proof}

We are now ready to prove \Cref{thm:intermediate_qcb}.
    \begin{proof}[Proof of \Cref{thm:intermediate_qcb}]
        We use $S_1,R,S$ to denote the intermediate random variables in \Cref{alg:quadratic_correlation_breaker} when applied on $(X,W,\alpha)$. Similarly, we define $S_1^{i},R^{i}$ to denote the intermediate random variables in the algorithm when applied to $(X,W^{i},\alpha^{i})$. 
        Let $\delta_1=2^{d_1}(2^{-s}+3\delta^{32}\cdot 2^{-4d_1})^{1/4}\leq O(\delta^8)$.
        From \Cref{thm:cauchy_schwarz_general} and the (1) our premise regarding $s$, and (2) the choice of $d_1,\lext$ in \Cref{alg:quadratic_correlation_breaker}, we have that with probability at least $1-\delta_1$ over $(s_1,z)\sim (S_1,Z)$, it holds that 
       \begin{equation}
       \label{eqn:cs_guarantee}
            R\approx_{\delta_1}U_{d_1}.
       \end{equation}

       Note that $R,R^{[t]}$ depend on both  $Y$ and $Z$. In the next step of the proof, we will show that we can fix additional information to make $R,R^{[t]}$ independent of $W$, while retaining the uniformity of $R$. Once we have this, we can use the definition of the advice correlation breaker (\Cref{defn:correlation_breaker}) to complete the proof. 

       We first fix $(S_1,S_1^{[t]})$. From the above argument, we have that with probability $1-\sqrt{\delta_1}$ over $(S_1,S_1^{[t]})$, it holds that $(R,Z)\approx_{\sqrt{\delta_1}}(U_{d_1},Z)$. Furthermore, $\hinf(W\mid S_1=s_1,S_1^{[t]}=s_1^{[t]})\geq d-td_0-O(\log(1/\delta))\geq d/2$ (from the choice of $d,d_0$ in \Cref{alg:quadratic_correlation_breaker}) with probability $1-O(\sqrt{\delta_1})$ over $(s_1,s_1^{[t]})\sim (S_1,S_1^{[t]})$ from \Cref{lem:chain_rule_smooth}. Thus these two events happen with probability at least $1-O(\sqrt{\delta_1})$ over $(S_1,S_1^{[t]})=(s_1,s_1^{[t]})$. We have $R=L(f(Y,Z))$ for some linear function $L$ (which depends on $s_1$). Similarly, we have $R^{i}=L^i(f(Y,Z))$ where $L^i$ is determined by $s_1^{i}$. Note that $Z$ is not uniform anymore.

       From \Cref{lem:partial_input_reduction}, we have a full rank linear map $M:\F_2^{s}\to \F_2^{s-\ell}$ with $\ell=d_1+O(\log (1/\delta))$ such that 
       \begin{equation}
       \label{eqn:input_reduce_cb}
       (R=L(f(Y,Z),Z,M(Y))\approx_{O(\sqrt{\delta_1})}(R=L(f(Y,Z)),Z',U_{s-\ell})
       \end{equation}
        for $U,Y,Z,Z'$ independent and $Z,Z'$ identically 
        distributed. 

        Let $e_1,\ldots, e_\ell$ be a basis for the nullspace of $M$ and $e_{\ell+1},\ldots, e_{s}$ be a completion of the basis to all of $\F_2^{s}$. We have that $Y=A+B$ for independent $A, B$ where (1) $A$ is uniform over the nullspace of $M$, and $B$ is uniform over the span of $\{e_{\ell+1}, \ldots, e_{s}\}$. Overloading notation, we have $A=A(P)$ and $B=B(Q)$ for linear maps $A$ and $B$ of appropriate dimension and independent uniform seeds $P=U_\ell$ and $Q=U_{s-\ell}$.
        
        Using the fact that $A,B$ are linear and overloading notation, we henceforth represent the source as $X=f(P,Q,Z)$. Now, rewriting \Cref{eqn:input_reduce_cb}, we have 
        \begin{equation}
        \label{eqn:input_reduce_cb_v2}
        (R, Z, M(B(Q)))\approx_{\delta_2}(R,Z', U_{s-\ell})
        \end{equation} where $\delta_2=O(\sqrt{\delta_1})$.
        This implies that $R$ is close to uniform even conditioned on $Q$ and $Z$. Now, we will argue that we can fix $Q$ and a few bits of information from $Z$ to make $(R,R^{[t]})$ independent of $(W,W^{[t]})$ while retaining the uniformity of $R$. Let $R=g(P,Q,Z)$ and $R^{i}=g^{i}(P,Q,Z)$ for quadratic maps $g,g^{[t]}$ each having output length $d_1$. 

        We have that 
        \[
        g(P,Q,Z)=g_1(P,Q)+g_2(Z)+g_3(P,Z)+g_4(Q,Z).
        \]
        Similarly, we have 
        \[
        g^i(P,Q,Z)=g_1^i(P,Q)+g_2^i(Z)+g_3^i(P,Z)+g_4^i(Q,Z).
        \]
        After fixing $Q$, we have that that with probability at least $1-\sqrt{\delta_2}$ over $q\sim Q$, it holds that
        \[
        (g(P,q,Z), Z)\approx_{\sqrt{\delta_2}}(g(P,q,Z),Z').
        \]
        Recall that $Z,W,W^{[t]}$ are independent of $P, Q$ and hence fixing $Q$ does not affect them. 
        Now, $g,g^{[t]}$ are only functions of $P,Z$. Observe that $\{g_3,g_3^{1},\ldots,g_3^{t}\}$ each depend on at most $d_1\ell$ one-bit linear functions of $Z$ as $P$ is uniform on $\ell$ bits. Let $(c(Z),c^{[t]}(Z))$ store these linear function in $(t+1)d_1\ell$ bits. Thus, fixing $(c(Z),c^{[t]}(Z),g_2(Z),g_2^{[t]}(Z),g_4(Z),g_4^{[t]}(Z))$ makes $R,R^{[t]}$ solely depend on $P$ while retaining uniformity of $R$ (for the latter guarantee, we use \Cref{eqn:input_reduce_cb_v2}). 

        We now fix $C(Z)=(c(Z),c^{[t]}(Z),g_2(Z),g_2^{[t]}(Z),g_4(Z),g_4^{[t]}(Z))$. With probability $1-O(\delta_2^{1/4})$ over $c\sim C(Z)$, we have that 
        \begin{align*}
        \hinf(W\mid (S_1,S_1^{[t]})=(s_1,s_1^{[t]}),C(Z)=c)&\geq d-td_0-3(t+1)d_1\ell-O(\log(1/\delta))\\
        &\geq t^5(m+\log a\log(n/\delta))^2.
        \end{align*}
        Also, from \Cref{eqn:input_reduce_cb_v2}, we have that with probability $1-\delta_2^{1/4}$ over $c\sim C(Z)$, it holds that 
        \[
        R\approx_{\delta_2^{1/4}} U_{d_1}.
        \]
    Also, we now have that $(R,R^{[t]})$ is independent of $(W,W^{[t]})$ as the former is only a function of $P,Q$ and the latter is a function only of $Z$ (which is independent of $P,Q$). We are now ready to invoke the guarantee of the advice correlation breaker. From \Cref{thm:li_correlation_breaker}, we have that 
    \[
    (CB(W,R,\alpha)\approx_{\delta+O(\delta_2^{1/4})},U_m)\mid\{CB(W^{i},R^{i},\alpha^i)\}_{i\in [t]}.
    \]
    Recall that $\delta_2=O(\sqrt{\delta_1})$ where $\delta_1=O(\delta^{8})$. Thus, we have that the final error $O(\delta)$.
    \end{proof}
    
We are now ready to complete the proof of \Cref{thm:quad_correlation_breaking}. 
    \begin{proof}[Proof of \Cref{thm:quad_correlation_breaking}]
        We prove the theorem by first fixing information to make the random variables $W,W^{[t]}$ depend only on a subset of the bits generating the source. We then use the $q$-purity of $X$ and \Cref{thm:structure_subseed} to decompose the original source into subsources with large derivative entropy. Finally, we apply \Cref{thm:intermediate_qcb}.

\paragraph{Quadratic Conditioning}
        To implement the first step (which we call "quadratic conditioning"), we use \Cref{lem:input_reduction_short}. We have a full rank linear map $L:\F_2^{k+\ell}\to \F_2^{k+\ell-h}$ with $h=d+O(\log(1/\delta))$ such that 
        \begin{equation}
        \label{eqn:input_reduce_qcb}
        (p(V), L(V))\approx_{\delta^2} (p(V),U_{k+\ell-h}).
        \end{equation}
        where $U_{k+\ell-p}$ is independent of $V$.
        
    Let $e_1,\ldots, e_h$ be a basis for the nullspace of $L$ and $e_{h+1},\ldots, e_{k+\ell}$ be a completion of the basis to all of $\F_2^{k+\ell}$. We have that $V=A+B$ for independent $A, B$ where (1) $A$ is uniform over the nullspace of $L$, and $B$ is uniform over the span of $\{e_{h+1}, \ldots, e_{k+\ell}\}$. Overloading notation, we have $A=A(Z)$ and $B=B(Y)$ for linear maps $A$ and $B$ of appropriate dimension and independent uniform seeds $Z=U_h$ and $Y=U_{k+\ell-h}$. We can represent our source as $X=f(M_1(Y)+M_2(Z))$ for linear maps $M_1, M_2$ of appropriate dimensions. Overloading notation, we can rewrite the polynomials $p,p^{[t]}$ as being quadratic functions of $Y,Z$. We have 
    \[
    p(Y,Z)=p_1(Y)+p_2(Z)+p_3(Y,Z)
    \] and similarly for each of the polynomials in $p^{[t]}$ (say $p^i(Y,Z)=p_1^i(Y)+p_2^{i}(Z)+p_3^{i}(Y,Z)$ for each $i\in [t]$). 
    
    From \Cref{eqn:input_reduce_qcb}, we also have that with probability $1-\delta$ over $y\sim Y$, it holds that $p(y,Z)\approx_{{\delta}}U_d$. We will now convert $p,p^{[t]}$ to functions solely depending on $Z$ by fixing all dependencies on $Y$.  Each of $p_3,p_3^{[t]}$ depend on $dh$ linear functions of $Y$. We will fix these $dht$ linear functions along with $p_1(Y),p_1^{[t]}(Y)$. Recall that $h=d+O(\log(1/\delta))$. We thus need to fix at most $b=O(dt(d+\log(1/\delta)))$ quadratic functions of $Y$. 
    
    Let $\ctpartition:\F_2^{k+\ell-h}\to \F_2^{k+\ell-h-s}$ be the function given by \Cref{thm:cohen-tal} (when the input functions are given by the $b$ quadratic functions we just considered) where $s=C(k+\ell-h)/b-b\geq Ck/(td^2)$ (for some constant $C$) and $\ctpartition^{-1}(c)$ is an affine subspace for all $c$. We fix $c\sim \ctpartition(Y)$. We have that with probability $1-\delta$ over $c\sim \ctpartition(Y)$, it holds that  \begin{enumerate}
        \item $Y'=(Y\mid \ctpartition(Y)=c)$ is uniform over an affine subspace of dimension $s$,
        \item $W,W^{[t]}$ are independent of $Y'$, 
        \item $W\approx_{O(\delta)}U_d$ and 
        \item $\hinf^{\delta^3}(f(M_1(Y')+M_2(Z))\geq k-(k+\ell-h-s)-O(\log(1/\delta))\geq C(k/(td^2))-td^2+h-O(\log(1/\delta))$ for constant $C$ (by \Cref{lem:chain_rule_smooth} and the bound on $s$).
    \end{enumerate}

    Observe that $f(M_1(Y')+M_2(Z))$ is a subsource of $X$ of the form $f(U_V)$ for affine subspace $V$. Thus, by \Cref{lem:purity_preserving}, this subsource is still $q$-pure.
    
    We overload notation and redefine $X, f, Y$ such that the source we have arrived at is $X=f(Y,Z)$ where $Y=U_s$, $Z=U_h$ . 
    We have that (1) $W,W^{[t]}$ are functions only of $Z$, (2) $\hinf^{\delta^{3}}(X)\geq s+h-O(\log(1/\delta))$ and (3) $X$ us $q$-pure. We are now ready to move to the next step where we invoke our structure theorem. 
    \paragraph{Structure Theorem}
    We now call \Cref{clry:pure_quadratic_structure}. From the end of the previous paragraph we have that $X$ satisfies the premise of \Cref{clry:pure_quadratic_structure} (as it is $q$-pure for $q\leq c\sqrt{k/(td^2)}$)  Thus, we have that $X$ is $2\delta$-close to a convex combination of subsources of the form 
      \[
      X'(Y',Z)=g_1(Y')+g_2(Y',Z)+g_3(Z)
      \]
      such that (1) $\hinf(\partial_a g_1(Y'))= \Omega(\sqrt{s})=  \Omega(d\sqrt{t})$ for all $a\neq 0$, (2) $\hinf^{2\delta^2}(X')=\Omega(td^2)$, (3) $Y'=U_{s'}$ with $s'\geq s/2$, and (4) $g_2$ is bilinear.

    Importantly, this decomposition left $Z$ and hence $W,W^{[t]}$ unaffected. We are now ready to finish the proof.
    \paragraph{Quadratic Correlation Breaking}
    Our source satisfies the premise of \Cref{thm:intermediate_qcb}. Applying \Cref{thm:intermediate_qcb} implies \Cref{eqn:qcb_guarantee_main} (we absorb the fact that $W\approx_{O(\delta)}U_d$ into the error). 
    \end{proof}

\section{Rank combiners}
\label{sec:rank_combiner}
The main goal of this section is to construct a linear seeded extractor for quadratic sources that has unusually low error on the good seeds. Note that for affine sources this is immediate, via \Cref{lem:rao_affine_lext} (as linear functions are either constant or have zero bias). For quadratics, this is no longer immediate, as one can have non-constant quadratic polynomials achieving nearly any bias (consider for example inner product on \(2b\) variables which has bias $2^{-b}$). 

Having unusually good error is important in the proof of the extractor as we want our good seeds to have error much smaller than the inverse of the total number of seeds. This is not possible for general sources as linear seeded extractors that work for arbitrary sources have seed length at least $\Omega(\log(1/\epsilon))$ (see Theorem~5 in \cite{Aggarwal2019ExtractorLB}).

In this section, we construct such extractors for quadratic sources with high derivative entropy (with the error going to zero as the derivative entropy increases). The extractor we construct is also a linear seeded extractor for affine sources. This suffices for our applications as we can always reduce to one of these two cases. After we defined the notion of a \textit{rank combiner} and proved that it implies such extractors, ChatGPT Astra gave us an explicit construction (\Cref{thm:explicit_rank_combiner_f2}) which implies the following.
\begin{theorem}[Good linear seeded extractors for quadratics over $\F_2$]
\label{clry:explicit_rank_combiner_extractor}
For any $d\geq \log^{40}n$, $\epsilon\geq 1/n^2$ and $m\leq d^{1/10}$, there exists an explicit function $\rcombext:\F_2^{n}\times [N]\to \F_2^{m}$ with $N=\poly(n)$ such that  
\begin{enumerate}
 \item $\rcombext$ is a $(\epsilon,2^{(m-r)/2})$-strong linear seeded extractor with $r\geq (d^{1/10}/\log^{3/2}n)$ for the class of quadratic sources $X=f(U_k)$ satisfying $\hinf(\partial_a f(\cdot))\geq d$ for all non-zero $a$.
    \item $\rcombext$ is a $(\epsilon,0)$-strong linear seeded extractor for affine sources with entropy at least $d$.
\end{enumerate}
\end{theorem}

To construct this object, we introduce a new pseudorandom object that we call a \textit{rank combiner}. 
\begin{definition}
\label{def:rank_combiner}
   A $(d,r,\epsilon)$-rank combiner over a field $\mathbb{F}$ is a collection of matrices $W_1,\ldots, W_N$ in $\mathbb{F}^{m\times n}$ such that for all  square matrices $G_1,\ldots, G_n$ satisfying  $\dim \{G_iy\}_{i\in [n]}\geq d$ for all $y\neq 0$, it holds with probability $1-\epsilon$ over $t\sim [N]$ that
    \begin{equation}
        \rank_{\mathbb{F}}\left(\sum_{i=1}^{n}(c^{\top}W_t )_{i}G_i\right)\geq r
    \end{equation}
    for all non-zero $c\in \mathbb{F}^{m}$. We call $m$ the output length and $N$ the size.  
\end{definition}
We now argue why the above definition implies a linear seeded extractor with unusually low error for all good seeds, when applied to a source with high derivative entropy. 
\begin{lemma}[Rank combiners are good linear seeded extractors for quadratics]
\label{thm:rank_combiners_imply_seeded_extractors}
Let $W_1,\ldots,W_N\in \F_2^{m\times n}$ be a $(d,r,\epsilon)$ rank combiner. Then, the function $L:\F_2^{n}\times [N]\to \F_2^{m}$ defined as $L(x,i)=W_ix$ satisfies the following properties: 
\begin{enumerate}
    \item $L$ is an $(\epsilon,2^{(m-r)/2})$-strong linear seeded extractor for the class of quadratic sources $X=f(U_k)$ satisfying $\hinf(\partial_a f(\cdot))\geq d$ for all non-zero $a$.
    \item $L$ is an $(\epsilon,0)$-strong linear seeded extractor for affine sources with entropy at least $d$.
\end{enumerate}
\end{lemma}
\begin{proof}
We first prove (1). Let $f(x)=(f_1(x),\ldots, f_n(x))$ for quadratic functions $f_1,\ldots f_n$. Let $f_i=x^{\top}M_ix+b_i^{\top}x$ for matrix $M\in \F_2^{k\times k}$ and vector $b\in \F_2^{k}$. For any non-zero $a$, the source $\partial_a f(U)$ is the affine source over the affine subspace $c_a+S_a$ for $c_a=((M_i^{\top}a+b_i))^{\top}a)_{i\in [n]}$ and $\dim(S_a)=\dim(\vecspan\{(M_i+M_i^{\top})a\}_{i\in [n]})$. The entropy guarantee on $\partial_a f(.)$ translates to the condition 
$\dim(S_a)\geq d$. Define $G_1,\ldots, G_n$ as the matrices $G_i\coloneq M_i+M_i^{\top}$. Thus, we have that $\dim\{G_i a\}_{i\in [n]}\geq d$ for all non-zero $a$. Thus, these matrices satisfy the premise of the rank combiner. 

The guarantee of the rank combiner gives us a set $T\subseteq [N]$ of size $(1-\epsilon)N$ such that for all $t\in T$, it holds that 
\[
\rank_{\F_2}(\sum_{i=1}^{n}(c^{\top}W_t)_iG_i)\geq r
\] for all non-zero $c$. 
We will argue that for all $t\in T$, holds that
\[
L(f(U),t)\approx_{2^{(m-r)/2}}U_m.
\]
Establishing the above inequality would complete the proof of $(1)$. Let $g_t(x)=L(f(x),t)$ be the quadratic map which we are analyzing. We will argue that for all non-zero $c\in \F_2^{m}$, it holds that $c^{\top}g_t(U_k)$ has bias at most $2^{-r/2}$. The final claim then follows from Vazirani's XOR lemma \Cref{fact:vazirani_xor}. We have that 
\[
c^{\top}g_t(x)=x^{\top}(\sum_{i=1}^{n}(c^{\top}W_t)_iM_i)x+(\sum_{i=1}^{n}(c^{\top}W_t)_ib_i^{\top})x
\]
We have that 
\[
(\E_U[(-1)^{c^{\top}g_t(U)}])^2\leq \E_{U,a}[(-1)^{\partial_a (c^{\top}g_t(U))}]
\]
where $a$ is uniform in $\F_2^{k}$. Observe that $\partial_a (c^{\top}g_t(U))$ is a linear function that is either constant or has bias zero. Thus, it suffices to bound the probability over $a$ that $\partial_a(c^{\top}g_t(.))$ is a constant. It is easy to see that this happens if and only if $(\sum_{i=1}^{n}(c^{\top}W_t)_iG_i)a=0$. From our rank combiner guarantee, this happens with probability $2^{-r}$ over $a\sim \F_2^{k}$. Thus, we have that 
\[
|\E_U[(-1)^{c^{\top}g_t(U)}]|\leq 2^{-r/2}.
\]
This completes the proof of (1).

We now prove the affine case (2). Let $X=f(U)$ satisfy the condition required by (1). For this step, a $(d,1,\epsilon)$ rank combiner suffices. An affine source $X$ of entropy $k\geq d$ can be be defined as the output of an affine map acting on $k$ uniform bits, that is $X=AU+b$ for uniform $U$ for $A\in \F_{2}^{n\times k}$ with $\rank(A)=k$. To prove the lemma statement, it suffices to show that with probability $1-\epsilon$ over $i\sim [N]$, it holds that the random variable $W_iAU$ is uniform. 

    Let $a_i$ be the $i$-th row of $A$. We also use $a_i$ to refer to the corresponding element of $\F_{2^k}$ when the vector is viewed as a field element. Let $M_i$ be the matrix such that $M_i x=a_i x$ when $x$ is appropriately interpreted as a vector and field element respectively. We have $\rank_{\F_2}(A)=\dim_{\F_2}\{a_i\}_{i\in [n]}=k$. Thus, for any non-zero $x$, we have $\dim_{\F_2}\{M_ix\}_{i\in [n]}=k\geq d$. Thus, $\{M_i\}_{i\in [n]}$ satisfies the premise of the rank combiner. Thus, with probability at least $1-\epsilon$ over $i\sim [N]$, it holds that $\rank(\sum_{j=1}^{n}(c^\top W_i)_{j}M_j)\geq 1$ for all $c\in \F_{2}^{m}$. Since $M_i$'s are denoting multiplication by field elements, we have that with probability $1-\epsilon$ over $i\sim [N]$, it holds that 
    \[
    \sum_{j=1}^{n}(c^\top W_i)_j a_j\neq 0    
    \] for all non-zero $c\in \F_{2}^{m}$. This proves that $c^{\top}W_iX=c^{\top}W_iA U+c^{\top}W_iA b$ is uniform for all non-zero $c$. Hence we have that $L$ is a $(d,\epsilon)$ strong linear seeded extractor for affine sources over $\F_2$. 
\end{proof}


Like many pseudorandom objects, the existence of great rank combiners is easy to show via the probabilistic method.

\begin{theorem}[Probabilistic existence of rank combiners]
\label{thm:rank_combiner_probabilistic}
   There exists a $(d,d-m-\log(1/\epsilon)-1,\epsilon)$-rank combiner over $\F_2$ with output length $m$ and size $\poly(nk^2/\epsilon)$.
\end{theorem}

We provide a proof in the appendix (\Cref{sec:additional_proof}).

\subsection{Rank combiners over large finite fields}
Our explicit construction of rank combiners proceeds in two steps. First, we give a construction over large fields of polynomial size. Then, we give a field size reduction argument using a linear seeded extractor in \Cref{sec:field_size_reduction}.
\label{sec:rank_combiner_large_field}

The proof of the main theorem of this section uses two additional lemmas (\Cref{thm:rank_combiner_polynomial_entries,lem:multiplicity_lower_bound}) which we prove immediately afterward.
\begin{theorem}
    \label{thm:rank_combiners_large_field}
     For any finite field $\F$ with $|\F|\geq 16mn^2/\epsilon$ and $m\leq \sqrt{d}/2$, there exists an explicit $(d,\sqrt{d}/(2m),\epsilon)$-rank combiner over $\mathbb{F}$ with output length $m$ and size $|\F|$. 
\end{theorem}
\begin{proof}
   We will first describe the construction before proving correctness. We will describe a matrix $W(\beta)\in \mathbb{F}^{m\times n}$ for $\beta\in \F$. The final rank combiner is $\{W(\beta)\}_{\beta\in \F}$.

   Choose an arbitrary set of distinct non-zero elements $\lambda_1,\ldots \lambda_{n}\in \mathbb{F}$. Let $w_1,\ldots,w_n$ be positive integers forming a Sidon set with $\max_{i}\{w_i\}\leq  8n^2$ (see \cite{erdos_sidon} for a construction). We define the entries of $W(\beta)$ as $(W(\beta))_{ji}=\lambda_i^{j-1}\beta^{w_i}$. We will now prove that that with probability at least $1-\epsilon$ over $\beta\sim \F$ we have 
\begin{equation}
    \label{eqn:good_b}
    \rank_{\mathbb{F}}\left(\sum_{i=1}^{n}(c^{\top}W(\beta))_{i}G_i\right)\geq \sqrt{d}/(2m)
\end{equation}
for all non-zero $c\in \F^{m}$. 
We say that $\beta\in \F$ is \textit{bad} if $W(\beta)$ does not satisfy \Cref{eqn:good_b} for some non-zero $c\in \F^{m}$. Let $B$ be the set of bad field elements. We want to show that $|B|\leq \epsilon |\F|$.

We now describe our high level plan. For any $\beta\in B$, there exists a non-zero vector $c_{\beta}\in \F^{m}$ such that \Cref{eqn:good_b} is not satisfied by $W(\beta)$. First we will construct a low degree polynomial map $c:\F\to \F^{m}$ such that $c(\beta)\in \vecspan\{c_\beta\}$ for all $\beta\in \F$ and $c(0)\neq 0$. This will give us a matrix with polynomial entries $M(\beta)=\sum_{i=1}^{n}(c(\beta)^{\top}W(\beta)))_{i}G_i$. We then use \Cref{thm:rank_combiner_polynomial_entries} to isolate a large sub-matrix with non-zero determinant. We then show that any bad $\beta$ that has low rank within this sub-matrix must have large multiplicity in the determinant polynomial. This allows us to bound $|B|$. 

First, we construct a polynomial map $c$ such that $c(\beta)\in \vecspan\{c_{\beta}\}$ for all $\beta\in B$ and $c(0)\neq 0$. We will show that there exists a polynomial $c$ of degree $\ell=|B|(1-1/m)$ that satisfies these properties. Finding such a $c$ is equivalent to solving the following homogeneous system of equations for the coefficients of $c$ in $\F^{m(\ell+1)}$ and $(\alpha_\beta)_{\beta\in B}\in \F^{|B|}$:
\[
c(\beta)=\alpha_\beta c_{\beta}, \text{ for all $\beta\in B$}.
\]

There are $m|B|$ constraints and the number of variables is $m(\ell+1)+|B|$ and the system is homogeneous. From our choice $\ell=|B|(1-1/m)$, system has more variables than constraints and is homogenous. Thus, there exists a non-zero solution. This solution must have $c\neq 0$ as $(\alpha_{\beta})_{\beta\in B}$ is forced to be $0$ when $c=0$. We can make $c(0)\neq 0$ by factoring out lowest degree term if our original solution had no constant term as this preserves the rank. Thus, we have obtained the polymial map $c$ that we desired. 

Now consider the matrix $M(\beta)=\sum_{i=1}^{n}(c(\beta)^{\top}W(\beta)))_{i}G_i$. We have that $\rank(M(\beta))\leq r$ for all $\beta\in B$ from (1) the fact that $c(\beta)=c_{\beta}$ or $c(\beta)=0$ for all $\beta \in B$ and (2) the definition of $c_\beta$. Expanding using the definition of $W(\beta)$, we obtain that 
\[
M(\beta)=\sum_{i=1}^{n}b_i(\beta)\beta^{w_i}G_i   
\]
where $b_i(\beta)=\sum_{j=1}^{m}c_j(\beta)\cdot \lambda_i^{j-1}$. Since $c(0)$ is non-zero, we have that $b_i(0)=\sum_{j=1}^{m}c_j(0)\cdot \lambda_{i}^{j-1}$ is only zero for at most $m$ indices as the $\lambda_i$'s are distinct. Thus, $M(\beta)$ satisfies the premise of \Cref{thm:rank_combiner_polynomial_entries} with $p=m$. Since $m\leq \sqrt{d}/2$, we have that $M(\beta)$ has a non-zero $\sqrt{d}\times \sqrt{d}$ minor. 

We now restrict to the $\sqrt{d}\times \sqrt{d}$ submatrix of $M(\beta)$ that has non-zero determinant. Call this submatrix $P(\beta)$. We will argue that $\rank(P(\beta))\geq \sqrt{d}/(2m)$ with probability $1-\epsilon$ over $\beta\sim \F$. For any $\beta\in B$, we have that $\rank(P(\beta))<\sqrt{d}/(2m)$ and hence from \Cref{lem:multiplicity_lower_bound}, all elements of $B$ have multiplicity at least $\sqrt{d}(1-1/(2m))$ in $\det(P(\beta))$. Recall that $\deg(c)=\ell\leq |B|(1-1/m)$. This implies that that $\deg(\det(P(\beta)))\leq (\deg(c)+\max_{i}\{w_i\})\sqrt{d}\leq \sqrt{d}(|B|(1-1/m)+8n^2)$ as the elements of the sidon set are at most $8n^2$.  Combining the multiplicity lower bound and the degree upper bound, we obtain $|B|\leq 16n^2m$. Since $|\F|\geq 16mn^2/\epsilon$, we have that the probability of sampling $\beta$ from $B$ is at most $\epsilon$. This completes the proof.
\end{proof}
We use the following lemma which considers constructions with polynomial entries.

\begin{lemma}
    \label{thm:rank_combiner_polynomial_entries}
   Let $G_1,\ldots,G_n\in \F^{k\times k}$ such that $\dim_{\F}\{G_iy\}\geq d$ for all non-zero $y\in \F$. Let $w_1,\ldots, w_n$ be a set of positive integers that form a Sidon set. Then, for any vector of polynomials $b(t)\in (\F[t])^{n}$ with $|\{i\mid b_i(0)\neq 0\}|\geq n-p$, it holds that the matrix
   \[
   M(t)=\sum_{i=1}^{n}b_i(t)t^{w_i}G_i 
   \]
   has a non-zero $r\times r$ minor with $r\geq \min(\sqrt{d},\frac{d}{2p})$.
\end{lemma}
\begin{proof}
    We will argue that the rank of $M(t)$ over the field of rational functions $\F(t)$ is at least $r$. We show that there every vector $g(t)\in (\F(t))^{k}$ in its kernel has at least $r$ non-zero entries. 
    
    For sake of contradiction assume that a sparse coefficient vector exists. Without loss of generality assume that $g(t)=(g_1(t),\ldots, g_r(t),0,\ldots 0)$ and that all the $g_i$'s are polynomials. Let $f_1,\ldots f_{\tilde{r}}$ be a basis for $\vecspan\{g_i\}_{i\in [r]}$ such that $\mindeg(f_i)\neq \mindeg(f_j)$ for $i\neq j$ (here $\mindeg(f)$ is the degree of the smallest non-zero monomial in $f$).
    
    Let $g(t)=\sum_{j=1}^{\tilde{r}}c_j f_j(t)$ where $c_j\in \F^{k}$. Thus, we have that
     \begin{equation}
        \label{eqn:rank_implies_zero}
        \sum_{j=1}^{\tilde{r}}\sum_{i=1}^{n}b_i(t)t^{w_i}f_j(t)G_i\cdot c_j=0.
     \end{equation}
    We want to use this equation to arrive at a contradiction. We will show that \begin{equation}
        \label{eqn:rank_bound_dependencies}
        \dim_{\mathbb{F}}\{c\in \F^{n\tilde{r}}\mid \sum_{i=1}^{n}\sum_{j=1}^{\tilde{r}}c_{ij}b_i(t)t^{w_i}f_j(t)=0\}\leq pr+\binom{r}{2}.
    \end{equation}
    We first show argue how \Cref{eqn:rank_bound_dependencies} implies a contradiction and then we will prove it. 
    Consider the matrix $B$ in $\F^{k\times n\tilde{r}}$ where the $(i,j)$-th column is $G_{i}c_j$ for $i\in [n], j\in [\tilde{r}]$. From \Cref{eqn:rank_implies_zero} and \Cref{eqn:rank_bound_dependencies}, we have that $\rank_{\F}(B)\leq pr+\binom{r}{2}$. Since $g$ is non-zero, there exists some $j$ such that $c_j$ is non zero. The matrix $[G_i c_j]_{i\in [n]}$ is a submatrix of $B$ and hence has rank at most $pr+\binom{r}{2}$ over $\F$.  From the premise of the lemma, we have $d\leq (pr+\binom{r}{2})\leq pr+r^2$. Solving for $r$, we get $r\geq \min(\sqrt{d},\frac{d}{2p})$.
    
    We now prove \Cref{eqn:rank_bound_dependencies}. To do this, we show that $\dim_{\F}\{b_i(t)t^{w_i}f_{j}(t)\}_{i\in [n],j\in [\tilde{r}]}\geq (n-p)\tilde{r}-\binom{\tilde{r}}{2}$. \Cref{eqn:rank_bound_dependencies} then follows by using the fact that $\tilde{r}\leq r$. Note that the dimension is lower bounded by the number of distinct min-degrees (degree of smallest non-zero monomial) among the $n\tilde{r}$ polynomials, as any set of polynomials with distinct min-degrees are linearly independent. Let $I$ be the set of indices $i$ for which $b_i(0)\neq 0$. The number of distinct min degrees is lower bounded by the terms only involving indices in $I$. Note that for such indices $i$, it holds that $\mindeg(b_i(t)t^{w_i}f_j(t))=w_i+\mindeg(f_j)$ for all $j \in [\tilde{r}]$. 
    
    Let $S_j=\{d\mid d=\mindeg(f_j(t)t^{w_i})\text{ for some $i\in I$}\}$. Clearly, the distinct min-degree count is lower bounded by $|\cup_{j=1}^{\tilde{r}}S_j|$. Furthermore, for any $i\neq j$, it holds that $|S_i\cap S_j|\leq 1$. To see this, note that for any $a,b\in [n]$, the equation $w_a+\mindeg(f_i)=w_b+\mindeg(f_j)$ fully determines the indices $a$ and $b$ (as $w_i$'s form a Sidon set). Thus, from inclusion-exclusion, we have that
    \[
    |\cup_j S_j|\geq (n-p)\tilde{r}-\binom{r}{2}.    
    \]This completes the proof of the dimension bound and hence the theorem.
\end{proof}

We need the following standard lemma about the determinant of matrices with polynomial entries. A similar statement and argument appears in \cite[Claim 16]{gk_subspace_designs}.
\begin{lemma}
    \label{lem:multiplicity_lower_bound}
    Let $P(t)\in (\F[t])^{k\times k}$ be a matrix with polynomial entries with non-zero determinant. Then any $\beta\in \F$ with $\rank_{\F}(P(\beta))\leq r$ is a root of $\det(P(t))$ with multiplicity at least $k-r$. 
\end{lemma}
\begin{proof}
    Take any $\beta$ satisfying $\rank(P(\beta))\leq r$. Let $P_i(t)$ be the $i$-column of $P$. We have that $P_i(t)=P_i(\beta)+(t-\beta)q_i(t)$ where $q_i$ is a polynomial, as $\beta$ is a root for $P_i(t)-P_i(\beta)$. Define the matrices $P_{S}(t)$ for $S\subseteq [k]$ as the matrix for which $(P_s(t))_i=P_i(\beta)$ if $i\in S$ and $(P_S(t))_i=q_i(t)$ if $i \notin S$. By expanding the determinint we have that $\det(P(t))=\sum_{S\subseteq [k]}(t-\beta)^{k-|S|}\det(P_S(t))$. Observe that $\det(P_S(t))=0$ for all $|S|\geq r+1$ as $P(\beta)$ has rank at most $r$ and $P_S(t)$ contains more than $r$ columns of $P(\beta)$. Thus, the only surviving terms in the expansion of $\det(P(t))$ are divided by $(t-\beta)^{k-r}$ and this completes the proof.
\end{proof}

\subsection{Field Size Reduction}
\label{sec:field_size_reduction}
The next lemma is used to translate dimension guarantees between $\F_2$ and $\F_{2^{h}}$.
\begin{lemma}
    \label{lem:rank_transfer_small_to_large}
   Let $G_1,\ldots, G_n \in \F_2^{k\times k}$ such that $\dim_{\mathbb{F}_2}\{G_iy\}_{i \in [n]}\geq d$ for all non-zero $y\in \mathbb{F}_2^k$. Then, for any positive integer $h$, it holds that $\dim_{\mathbb{F}_{2^h}}\{G_iz\}_{i\in [n]}\geq d/h$ for all non-zero $z\in \mathbb{F}_{2^h}^k$. 
\end{lemma}
\begin{proof}
    We will prove the contrapositive. Suppose there is a non-zero $z\in \F_{2^h}$ such that $\dim_{\mathbb{F}_{2^h}}\{G_iz\}< d/h$. Let $e_1,\ldots,e_{h}$ be a basis of $\F_{2^{h}}$ over $\F_2$. Let $V=\vecspan_{\F_{2^h}}\{G_i z\}$. First observe that $\dim_{\F_2}(V)=h\cdot \dim_{\F_{2^h}}(V)< d$ (the first equality holds because $\{e_iv_j\}_{i\in[j], j\in [\dim(V)]}$ is a basis of $V$ over $\F_2$ whenever $v_1,\ldots, v_{\dim(V)}$ is a basis of $V$ over $\F_{2^h}$). Let $z=\sum_{i=1}^{h}c_ie_i$ where $c_i\in \F_2^{k}$. Since $z\neq 0$, there exists $j$ such that $c_j\neq 0$. Now consider the function $\pi$ that maps $G_i z$ to $G_i c_j$. Extend this map linearly over all of $V$. Clearly $\pi(V)$ is a subspace and $\dim_{\F_2}(\pi(V))\leq \dim_{\F_2}(V)< d$. This implies that $\dim_{\F_2}\{G_i c_j\}<d$. This completes the proof of the contrapositive. 
\end{proof}

We are now ready to state and prove our theorem on field size reduction.
\begin{theorem}[Field size reduction for rank combiners]
\label{thm:rank_combiner_field_size_reduction}
  Let $U_1,\ldots,U_N\in \F_{2^{h}}^{m\times n}$ be an explicit $(d,r,\epsilon)$-rank combiner over $\F_{2^h}$ and let $\lext:\F_2^{hm}\times \cube{\ell}\to \F_2^{m'}$ be an explicit $(m,\delta)$-strong linear seeded extractor . Then, there is an explicit $(dh,r/(hm),\epsilon+\delta)$-rank combiner $W_1,\ldots, W_{M=N\cdot 2^\ell}$ in $\bcube{m'\times n}$.
\end{theorem}
\begin{proof}
Let $G_1,\ldots, G_n$ be matrices in $\F_2^{k\times k}$ satisfying $\dim_{\F_2}\{G_iy\}_{i\in [n]}\geq dh$ for all non-zero $y\in \F_2^{k}$.
Let $e_1,\ldots, e_{h}$ be basis vectors for $\mathbb{F}_{2^{h}}$ when it is treated as a vector space over $\mathbb{F}_2$. Note that for a fixed seed~$s$,  the function computed by $\lext(\cdot,s)$ can be denoted by a matrix $L_s$ in $\bcube{m'\times  hm}$.
 For $U\in \mathbb{F}_{2^{h}}^{m\times n}$, write $U=\sum_{j=1}^{h}M_{j}(U)e_j$ with $M_j(U)\in \F_2^{m\times n}$ and define 
\[
M(U) := \begin{pmatrix}
M_1(U)\\
\vdots\\
M_h(U)
\end{pmatrix},
\quad
W(U,s) := L_s M(U),
\quad
W(i,s) := W(U_i,s).
\]
We will show that $\{W_{(i,s)}\}_{i\in [N], s\in \bcube{\ell}}$ is a $(dh,r/(hm),\epsilon+\delta)$-rank combiner over $\F_2$.
  The key property of the linear seeded extractor we use is that the set of vectors given by its seeds evade small subspaces. From \Cref{lem:rank_transfer_small_to_large}, we have that $\rank_{\F_{2^h}}\{G_iz\}\geq d$ for all non-zero $z\in \F_{2^h}^{k}$. Then, from the definition of a rank combiner, we have that with probability $1-\epsilon$ over $j\sim [N]$, it holds that $\rank_{\mathbb{F}_{2^h}}\left(\sum_{i=1}^{n}(c^{\top}U_j)_iG_i\right)\geq r$
  for all non-zero $c\in \F_{2^h}^{m}$. We set $U=U_j$ for $U_j$ that satisfied the previous condition. We use $M$ to denote $M(U)$ as defined above. 
    Let $V\in \bcube{hm}$ be the span (over $\mathbb{F}_2$) of vectors $v$ satisfying
    \begin{equation}
    \label{eqn:bad_weight}
        \rank_{\mathbb{F}_2}\left(\sum_{i=1}^{n}(v^\top M(U))_iG_i\right)< r/(hm).
    \end{equation}
    \begin{claim}
    \label{claim:dim_upper_bound_bad}
        $\dim_{\mathbb{F}_2}(V)\leq hm-m$.
    \end{claim}
    \begin{proof}
    We will argue that $\dim_{\mathbb{F}_2}(V)\leq hm-m$. Take a basis $v_1,\ldots, v_t$ of $V$ such that each $v_i$ satisfies \Cref{eqn:bad_weight}. Let $\tilde{V}$ be the subspace over $\mathbb{F}_{2^{h}}^{hm}$ obtained by taking the span of $v_1,\ldots v_t$ over $\mathbb{F}_{2^{h}}$. Clearly, $\dim_{\mathbb{F}_2}(V)=\dim _{\mathbb{F}_{2^h}}(\tilde{V})$ as linear independence of the basis vectors is preserved. Thus, it suffices to bound the latter quantity. 
    For any $\tilde{v}\in \tilde{V}$, we have that
    \begin{equation*}
        \rank_{\mathbb{F}_{2^h}}\left(\sum_{i=1}^{n} (\tilde{v}^\top M)_iG_i\right)\leq \sum_{j=1}^{t}\rank_{\mathbb{F}_{2^h}}\left(\sum_{i=1}^{n} (v_{j}^\top M)_iG_i\right)
    \leq \sum_{j=1}^{t}\rank_{\mathbb{F}_{2}}\left(\sum_{i=1}^{n} (v_{j}^\top M)_iG_i\right)<r
    \end{equation*}
    where the first inequality follows by subadditivity of rank and the second by the fact that rank only increases when looking at a subfield. The final inequality follows from \Cref{eqn:bad_weight} and the fact that $t\leq hm$. 

    For any $0\neq c=(c_1,\ldots,c_m)\in \mathbb{F}_{2^h}^{m}$, its holds that 
    $\rank_{\mathbb{F}_{2^h}}\left(\sum_{i=1}^{n}(c^\top U)_iG_i\right)\geq r$. Expanding using basis elements, we have that $\rank_{\mathbb{F}_{2^h}}\left(\sum_{i=1}^{n}\sum_{j=1}^{h}e_j(c^\top M_j)_iG_i\right)\geq r$. Defining $v(c)=(e_ic_j)_{i\in [h],j\in [m]}$, we have that $\rank_{\mathbb{F}_{2^h}}\left(\sum_{i=1}^{n}(v(c)^\top M)_iG_i\right)\geq r$ for all $c\neq 0$. This means that for all $c\neq 0$, it holds that $v(c)\notin \tilde{V}$. Clearly the map $c\mapsto v(c)$ is a linear injective map. Thus the subspace $\overline{V}=\{v(c)\mid c\in \mathbb{F}_{2^h}^m\}$ satisfies $\dim_{\mathbb{F}_{2^h}}(\overline{V})= m$ and $\tilde{V}\cap \overline{V}=\{0\}$. This implies that $\dim_{\mathbb{F}_{2^h}}(\tilde{V})=\dim_{\mathbb{F}_2}(V)\leq hm-m$.  
    \end{proof}

We are now ready to complete the proof. Define $V^{\perp}=\{u\in \mathbb{F}_2^{hm}\mid u^\top v=0 \text{ for all $v\in V$}\}$. From \Cref{claim:dim_upper_bound_bad}, we have that $\dim_{\mathbb{F}_2}(V^{\perp})\geq m$. Let $X$ be a uniform random variable over $V^{\perp}$. From the fact that $\lext$ is an $(m,\delta)$-strong linear seeded extractor and \Cref{lem:rao_affine_lext}, we have  with probability $1-\delta$ over $s\in \bcube{\ell}$, that $b^{\top}\lext(X,s)$ is a uniform bit for all non-zero $b\in \mathbb{F}_2^{m'}$. However, this must mean that $b^{\top}L_s\notin V$ for all non-zero $b$ as otherwise $b^{\top}\lext(X,s)=0$ by the definition of $V^{\perp}$. Since $b^{\top}L_s\notin V$ for all non-zero $b$, it implies that $W=L_s\cdot M$ satisfies \[
    \rank_{\mathbb{F}_2}\left(\sum_{i=1}^{n}(b^{\top}W )_{i}G_i\right)\geq r/(hm).\]

    Thus, we have shown that for each good matrix $U_i$ in the original rank combiner, $1-\delta$ fraction of the seeds $s\in \F_{2}^\ell$ lead to a good matrix $W(U_i,s)$. Thus, the overall success probability is $1-(\epsilon+\delta)$ and the proof is complete. 
\end{proof}
\subsection{Putting Things Together}
\label{sec:rank_combiner_final}
\begin{theorem}[Explicit rank combiner over $\F_2$]
\label{thm:explicit_rank_combiner_f2}
    For any $d\geq \log^{40}n$ and $\epsilon\geq 1/n^2$, there exists an explicit $(d,(d^{1/10}/\log^{3/2}n),\epsilon)$-rank combiner over $\mathbb{F}_2$  with output length $d^{1/10}$ and size $\poly(n)$.
\end{theorem}
\begin{proof}
 Take $h=O(\log(n))$ and $m=d^{1/5}/2$. Let $U_1,\ldots, U_N\in \F_{2^h}^{m\times n}$ be the $(d/h,d^{3/10}/\sqrt{h},\epsilon/2)$ rank combiner from \Cref{thm:rank_combiners_large_field} with $N=\poly(n)$ (we use the fact that $\epsilon\geq 1/n^2$). We will now use an appropriately chosen strong linear seeded extractor over $\F_2$ and apply \Cref{thm:rank_combiner_field_size_reduction}.

 We use the linear seeded extractor due to Li \Cref{thm:li_lext} with a small modification. Recall that the linear seeded extractor needed by \Cref{thm:rank_combiner_field_size_reduction} acts on $hm$ bits. Directly applying \Cref{thm:li_lext} with input length $hm$ forces the error to be at least $1/(hm)^2$. To overcome this, we modify Li's extractor as follows: (1) first pad the $hm$ length input string with zero's to make the total length $2n$, (2) apply the extractor from \Cref{thm:li_lext} with input length $2n$. As long as $m\geq \Omega(\log^8 n)$, the output length is $\sqrt{m}$ with error $1/(4n^2)\leq \epsilon/2$. From our assumption that $d\geq \log^{40} n$, we have that $m=d^{1/5}/2\geq \log^8 n$. Thus, \Cref{thm:rank_combiner_field_size_reduction} implies an explicit $(d,(d^{1/10}/\log^{3/2}n),\epsilon)$-rank combiner with output length $d^{1/10}$ and size $\poly(n)$.
 \end{proof}

This immediately gives the construction of the linear seeded extractor where the good rows have unusually good error. 
\begin{proof}[Proof of \Cref{clry:explicit_rank_combiner_extractor}]
    Truncate the output length of the rank combiner from \Cref{thm:explicit_rank_combiner_f2} and apply \Cref{thm:rank_combiners_imply_seeded_extractors}.
\end{proof}
\subsection{Provenance of the rank combiner construction}
\label{sec:provenance}
Related constructions and arguments have appeared in prior work studying rank condensers, subspace designs, and polynomial identity testing. In particular, constructions similar to the large field construction (\Cref{thm:rank_combiners_large_field}) appear in \cite{FS12,FSS14} (see Lemma~3.10 and Remark~3.11 in the full version of \cite{FSS14}). However, their constructions do not use the Sidon set weights which are used in the proof of \Cref{thm:rank_combiner_polynomial_entries}. Other papers using similar constructions include \cite{klivans_spielman} (see Section 3 of the full version) and \cite{fenner2016bipartite} (see Section 3 of the full version) in the context of polynomial identity testing and bipartite matching. The multiplicity argument for the determinant polynomial in \Cref{lem:multiplicity_lower_bound} appears in \cite{gk_subspace_designs} (see Claim~16). Recent work of \cite{GRSZ26} (Lemma~5.6) studied the related problem of symbolic determinant identity testing when restricted to the setting of rank one summands.
\section{Quadratic extractors}

\begin{algorithm}
    \caption{Quadratic Extractor ($\qext$)}
    \label{alg:quadratic_extractor}
   \begin{algorithmic}
    \STATE \textbf{Input: } $x\in \{0,1\}^{n}$--- an $n$-bit string.

    \STATE \textbf{Output:} $v\in \{0,1\}^m$--- output string.
    \STATE \hrulefill
   \STATE  \textbf{Sub-Routines and Parameters:}
   \STATE Let $C_1,C_2,C_3$ be appropriately chosen constants. Let $N=n^{C_1}$.
   \STATE Let $\rcombext:\bcube{n}\times [N]\to \F_2^{d}$ be the linear seeded extractor from \Cref{clry:explicit_rank_combiner_extractor}.
   \STATE Let $\qcb:\bcube{n}\times \bcube{d}\times [N]\to \{0,1\}$ be the quadratic correlation breaker from \Cref{thm:quad_correlation_breaking} with output length $1$, $t=(m\log n)^{C_2}$, $\delta=N^{-10t}$ and $d=t^5(\log (n/\delta))^3$. 
   \STATE Let $\nobfext:\{0,1\}^{N}\to \{0,1\}^m$ be the NOBF extractor from \Cref{thm:nobf_extractor}.
   \STATE \hrulefill
    \begin{enumerate}
        \item Let $w=\{\rcombext(x,i)\}_{i\in [N]}$.
        \item Let $s=\{\qcb(x,w_i,i)\}_{i\in [N]}$.
        \item Output $\nobfext(s)$.
    \end{enumerate}
\end{algorithmic}    
\end{algorithm}

We execute a win-win argument where we analyze the extractor differently based on if it contains a large subspace or not. We use the following elementary lemma from \cite{cohen-tal} that states that any quadratic map that is affine on an affine subspace $V$ is also affine on shifts of $V$. 
\begin{lemma}[Claim~3.2 from \cite{cohen-tal}]
\label{lem:ct_affine_restriction}
    Let $f$ be a quadratic map on $\F_2^{n}$ such that $f|_V$ is affine on an affine subspace $V\subseteq \F_2^n$. Then $f|_{u+V}$ is also affine, for all $u\in \F_2^{n}$.
\end{lemma}
The following lemma is now immediate.
\begin{lemma}
\label{lem:affine_tiling}
    Let $X=f(U)$ be a quadratic source such that $\hinf^{\epsilon}(X)\geq k$ and $f:\F_2^{k+\ell}\to \F_2^{n}$ for $\ell=O(\log(1/\epsilon))$. Suppose $X$ is not $q$-pure. Then, $X$ can be decomposed is $O(\epsilon)$ close to a  convex combination of affine sources with entropy $q-O(\log(1/\epsilon))$.
\end{lemma}
\begin{proof}
   From the fact that $X$ is not $q$-pure, we have an affine subspace $V=v+S$ for subspace $S$ such that $f|_{V}$ is affine and $\dim(f(V))\geq q$. Let $\dim(S)=d\geq q$. From \Cref{lem:ct_affine_restriction}, we have that $f|_{u+V}$ is affine for all shifts $u$. Let $C:\F_2^{k+\ell}\to \F_2^{k+\ell-d}$ be the full rank linear map with kernel $S$. $C^{-1}(c)$ is a shift of $V$ for all $c$. Thus, we have that $f(U)\mid \{C(U)=c\}$ is an affine source for all $c$. Also, $\hinf^{1/2}(f(U)\mid C(U)=c)\geq q-O(\log(1/\epsilon)$ with probability $1-O(\epsilon)$ over $c\sim C(U)$. Since affine sources are flat, \Cref{lem:smooth_entropy_to_entropy} implies that $\hinf(X\mid C(U)=c)\geq q-O(\log(1/\epsilon))$ with probability at least $1-O(\epsilon)$ over $c\sim C(U)$.
\end{proof}

\subsection{$\qext$ is an affine extractor}
First, we argue that the construction in \Cref{alg:quadratic_extractor} is an affine extractor. Similar constructions have been used in prior work (with $\rcombext$ replaced by a standard linear seeded extractor) and showing that $\qext$ is an affine extractor is relatively straightforward. 
\begin{theorem}
    \label{thm:qext_is_affine_extractor}
    There exists a constant $C>0$ such that for any affine source $X$ with $\hinf(X)=k\geq \log^C n$, it holds that $\qext(X)\approx_{n^{-\Omega(1)}} U_{k^{\Omega(1)}}$.
\end{theorem}
We defer the proof to the appendix \Cref{sec:qext_is_affine_extractor}.
\subsection{$\qext$ is a quadratic extractor}
We now state and prove the main theorem of the paper. 
\begin{theorem}
\label{thm:main_theorem_quadratic}
    There exists a constant $C>0$ such that for any quadratic source $X$ satisfying $\hinf(X)=k\geq \log^C n$, it holds that $\qext(X)\approx_{n^{-\Omega(1)}}U_{k^{\Omega(1)}}$.
\end{theorem}
\begin{proof}
We will first instantiate various parameters that we use throughout the proof. Let $m$ be the output length of the extractor with $m=k^c$ for some constant $c$ that we will choose later. We instantiate the linear seeded extractor $\rcombext$ from \Cref{clry:explicit_rank_combiner_extractor} with $\epsilon=1/n^2$ and derivative entropy requirement $k^{1/4}$ (we will choose the output length soon). Let $N=n^{C_1}$ be the number of seeds required by \Cref{clry:explicit_rank_combiner_extractor}. Let $t=(m\log n)^{C_2}$ be chosen such that \Cref{thm:nobf_extractor} gives a $(q=N/n^2,t)$-NOBF extractor with error $n^{-C_3}$ and output length $m$. We define an error parameter $\delta=N^{-10t}$ which will be the error of each almost-uniform bit in the NOBF source that we finally obtain. Let $d=C_4t^{5}(\log n\log(n/\delta))^2$ be the seed length of $\qcb$ when we have target error $\delta$ and $t$ as chosen earlier. $d$ will also be the output length of $\rcombext$ (we will analyze it's error later). 

We choose the constant $c$ in $m=k^{c}$ sufficiently small and the constant $C$ in $k\geq \log^C n$ sufficiently large such that 
\[
\max(t,d,\log(1/\delta))\leq k^{1/100}
\]
and $k^{1/3}$ is asymptotically greater than the entropy requirement of \Cref{thm:qext_is_affine_extractor}.

Our first step in the proof is to apply the input reduction lemma (\Cref{lem:input_reduction_entropy}): $X$ is $\delta$-close to a convex combination of  quadratic sources $X'$ with $\hinf^{\delta^6}(X')\geq k-1$ and $X'=f(U_{k+\ell})$ for quadratic $f$ and $\ell=O(\log(1/\delta))$. We will argue that $\qext$ extracts from any source $X'$ satisfying this property. Henceforth, without loss of generality, we assume that our source $X=f(U_{l+\ell})$ satisfies (1) $\hinf^{\delta^6}(X)\geq k-1$ and (2) $\ell=O(\log(1/\delta))$.

We now split the proof into two cases depending on if $X$ is $(ck^{1/3})$-pure for some small constant $c>0$.

\paragraph{Not pure case} 
Suppose $X$ is not $(ck^{1/3})$-pure. Then, \Cref{lem:affine_tiling} implies that $X$ is $O(\delta)$ close to a convex combination of affine sources with entropy $ck^{1/3}/2$ for constant $c'<c$. The theorem statement now follows from \Cref{thm:qext_is_affine_extractor}.

\paragraph{Pure case}
Suppose $X$ is $(ck^{1/3})$-pure. We now apply the structure theorem. \Cref{clry:pure_quadratic_structure} (with $s=k+\ell$) implies that $X$ is $\delta^2$ close to a convex combination of sources $X'$ satisfying (1) $X'=g(Y)$ for $Y=U_s'$ and $s'\geq s/2$, (2) $\hinf(\partial_a g(.))\geq k^{2/3}/(4c)$, (3) $\hinf^{\delta^4}(X)\geq s'-\ell$. From \Cref{lem:purity_preserving}, we have that each $x$' is still $(ck^{1/3})$-pure. We rename $X'$ as $X$ and $g$ as $f$ and henceforth consider $X$ satisfying the above property. 

Henceforth let $X=f(Y)$ where (1) $\hinf^{\delta^4}(X)=\overline{k}=\Omega(k)$, (2) $f:\F_2^{\overline{k}+\ell}\to \F_2^{n}$ is quadratic, and (4) $Y=U_{\overline{k}+\ell}$, (3) $\hinf(\partial_a f(Y))\geq \Omega(k^{2/3})$ for all non-zero $a$.
We will first argue that the random variable $W=(\rcombext(X,i))_{i\in [N]}$ has the following property: $W_i\approx_{\delta^2} U_d$ with probability $1-1/n^2$ over $i\sim [N]$. Recall the output length  $d$ of $\rcombext$ that was chosen earlier satisfies $d\leq k^{1/100}$. Thus, from our derivative entropy guarantee of $\Omega(k^{2/3})$, we have that $W_i\approx_{2^{-k^{1/40}}} U_d$ (with probability $1-1/n^2$ over $i\sim [N]$) and this error is much smaller than $\delta^2$. We also have that $W_i=p^i(Y)$ for some quadratic polynomial $p^i$.

Let $Q\subseteq [N]$ be the set of indices where $W_{i}\approx_{\delta^2} U_d$. Consider the random variable \[S=(\qcb(X,W^i,i))_{i\in [N]}.\] We will argue that for all sets $T\subseteq Q$ with $|T|\leq t$, we have that $|\E[\chi_T(S)]|\leq O(\delta)$.
This is immediate from \Cref{thm:quad_correlation_breaking} and the fact that $X,W$ satisfy the premise (by our choice of parameters) and the indices $i\in [N]$ form valid advice strings. 

Now, applying \Cref{fact:agm} implies that $S$ is $N^{-t}$-close to a $(q,t)$-NOBF source with $q=N/n^2$ and $t=(m\log n)^{C_2}$ chosen earlier. The theorem statement now follows from \Cref{thm:nobf_extractor}.
\end{proof}

\section{Applications}
We now state and prove improved extractors for $\nc$ and $\ac$ sources. Prior constructions for these sources were obtained by reducing to $1$-local sources (which are a special type of affine source). We give a more efficient reduction to $2$-local sources (which are a special type of quadratic source), and use this to obtain the improved results.

Before we show our results, we define an intermediate notion that is useful.
\begin{definition}[$d$-local NOBF source]
    A source $X$ is $d$ local NOBF if $X=f(U_s)$ for a $d$-local map $f$ such that for any $i\in [s]$, there exist $j(i)\in [n]$ such that $f_{j(i)}(x)=x_{i}$.
\end{definition}
Informally, a $d$-local NOBF source is a $d$-local source where the generating bits are printed out somewhere in the source.

We use the following theorem (from \cite{alrabiah_et_al:LIPIcs.ICALP.2022.10}) that reduces extracting from $d$-local sources to extracting from $d$-local NOBF sources. 

\begin{theorem}[Theorem~2 of \cite{alrabiah_et_al:LIPIcs.ICALP.2022.10}]
\label{thm:reduce_local_to_nobf}
    There exists a constant $c>0$ such that any $d$-local source $X$ with $\hinf(X)\geq k$ is $\epsilon$-close to a convex combination of $d$-local NOBF sources with min-entropy at least $k'$ for $k'\geq ck/(2^dd^2)$ and $\epsilon=2^{-ck'}$.
\end{theorem}

We first show that every $d$-local source of sufficient entropy is close to a convex combination of $2$-local sources.
\begin{theorem}
\label{thm:d_local_2_local}
     There exists a constant $c,C>0$ such that any $d$-local source $X$ with $\hinf(X)\geq k$ is $\epsilon$-close to a convex combination of $2$-local sources with min-entropy at least $k'$ for $k'\geq Ck^{3/2}/(2^{2d}\sqrt{n})$ and $\epsilon=2^{-ck/(2^dd^2)}$.
\end{theorem}
\begin{proof}
    From \Cref{thm:reduce_local_to_nobf} we have that $X$ is $2^{-ck'}$ close to a convex combination of $d$-local NOBF sources with min-entropy $k'\geq ck/(2^{d}d^2)$ for some constant $c>0$. Let $S_i\subseteq[k']$ be the set of input generating bits that the function $f_i$ depends on. Consider the set system $\mathcal{S}=\{T\mid\text { $T\subseteq S_i$ for some $i\in [n]$ and $|T|=3$}\}$. Clearly, we have $|\mathcal{S}|\leq n\cdot \binom{d}{3}$ as each $S_i$ has size at most $d$. Recall Turan's theorem for hypergraphs \cite{SPENCER1972183} which states that the independence number of a $3$-uniform hypergraph with $p$ vertices and $q$ edges is at least $C\min(p,p^{3/2}/\sqrt{q})$ for some constant $C$. Interpreting $\mathcal{S}$ as a hypergraph then implies that $\mathcal{S}$ has an independent set of size at least $C'(k')^{3/2}/\sqrt{nd^3}$ for some constant $C'$. 

    Let $I\subseteq [k']$ be the independent set of $\mathcal{S}$. From the definition of independent set, we have that $I$ does not contain $S_i$ for any $i\in [n]$. This means that $|I\cap S_i|\leq 2$ for all $i\in [n]$. Consider the source $X'$ obtained by conditioning on the generating bits in $[k']\setminus I$. By the intersection property described above, this source is $2$-local NOBF. Also, the min-entropy of the source is at least $|I|$ as the it is still NOBF (the unfixed generating bits are still printed out in the source). Thus, we have that our original source is $\epsilon=2^{-ck'}$ to a convex combination of $2$-local NOBF sources with entropy at least $|I|\geq C''k^{3/2}/(2^{2d}\sqrt{n})$ for some constant $C''>0$.
\end{proof}

\begin{theorem}[Extractors for $\nc$-sources]
\label{thm:ext_nc0}
 For any constant $\alpha>0$ and any $\nc$-source $X$ with $\hinf(X)=k\geq n^{1/3+\alpha}$, it holds that $\qext(X)\approx_{n^{-\Omega(1)}}U_{k^{\Omega(1)}}$
\end{theorem}
\begin{proof}
    The theorem follows from applying \Cref{thm:d_local_2_local}, \Cref{thm:main_theorem_quadratic} and the fact that a $2$-local source is a quadratic source.
\end{proof}

The following theorem by \cite{viola_circuit_sources} shows that $\ac$-sources can be reduced to $\nc$-sources (we instantiate the theorem to entropy close to $n^{1/3}$.
\begin{theorem}[\cite{viola_circuit_sources}, see Theorem~8.17 from \cite{goodman2023seedless}]
\label{thm:ac_0_to_nc_0}
    For any $\alpha>0$ and $\ac$-source $X$ satisfying $\hinf(X)\geq n^{1/3+\alpha}$, there exists constants $d,\beta$  such that $X$ is $n^{-\Omega(1)}$-close to a convex combination of $d$-local sources with entropy at least $n^{1/3+\beta}$.
\end{theorem}

Our theorem on extraction from $\ac$ sources is now immedaite. 
\begin{theorem}[Extractors for $\ac$-sources]
\label{thm:ext_ac0}
    For any constant $\alpha>0$ and any $\ac$-source $X$ with $\hinf(X)=k\geq n^{1/3+\alpha}$, it holds that $\qext(X)\approx_{n^{-\Omega(1)}}U_{k^{\Omega(1)}}$
\end{theorem}
\begin{proof}
    Apply \Cref{thm:ac_0_to_nc_0} and \Cref{thm:ext_nc0}.
\end{proof}
\section{Open Directions}
The natural open question is to construct better extractors for higher degree sources. We now discuss the parts of our proof that extend and the parts that need new ideas. 

The proof of quadratic conditioning \Cref{lem:quadratic_conditioning_technical_overview} naturally extends to all constant degree polynomials, as the lemma from \cite{cohen-tal} also has constant degree extensions. The proofs using Cauchy Schwarz in \Cref{thm:cauchy_schwarz_general} and \Cref{thm:intermediate_qcb} also extends to degree $d$ sources if we assume the premise that the sources corresponding to $(d-1)$-th directional derivatives have high entropy. 

The proofs of \Cref{thm:structure_injective,thm:structure_subseed} do not seem to extend immediately to degree $3$ and higher. A natural question is the following: can all degree $d$ sources with a min-entropy guarantee be decomposed into a convex combination of sources of one of two types: (1) degree $(d-1)$ sources with high entropy, or (2) sources with high $(d-1)$-th derivative entropy? Answering this question positively for cubics would already give an improved disperser by adapting our proof in \Cref{sec:disperser}. 

Another open direction is to find tensor analogues of rank combiners (\Cref{def:rank_combiner}). These seem useful to construct good linear seeded extractors for higher degree polynomial sources. 

Finally, an interesting question is to see if any of our techniques are useful for sources of the type $f(X+Y)$, where $X$ and $Y$ are independent sources each having a min-entropy guarantee, and $f$ is a quadratic map. This is the natural generalization of sum-set sources to higher degree.

\newpage
\bibliographystyle{alpha}
\bibliography{refs}
\newpage
\appendix

\section{High-entropy dispersers using the challenge-response framework}
\label{sec:disperser}
We now present an alternative construction that gives a disperser for quadratic sources with min-entropy $n-n^{0.1}$. This is a worse result than our main theorem (which gives an extractor at polylogarithmic entropy), however we believe the usage of the challenge-response framework in the proof is interesting in its own right. 
\begin{theorem}
\label{thm:main_disp}
    There exists an explicit function $\qdisp:\bcube{n}\to \bcube{}$ (\Cref{alg:quadratic_disperser}) that is a disperser for quadratic sources with entropy $n-n^{0.1}$. 
\end{theorem}

\begin{algorithm}
    \caption{Quadratic Disperser ($\mathrm{QDisp}$)}
    \label{alg:quadratic_disperser}
   \begin{algorithmic}
    \STATE \textbf{Input: } $x\in \{0,1\}^{n}$--- an $n$-bit string.

    \STATE \textbf{Output:} $v\in \{0,1\}$--- output bit.
    \STATE \hrulefill
   \STATE  \textbf{Sub-Routines and Parameters:}
   \STATE Let $\Delta=n^{0.1}$, $\ell=0.1n^{0.1},  d_1=100\Delta$ , $d_2=O(\log n),\epsilon=2^{-10n^{0.1}}$.
   \STATE Let $\cglext:\bcube{n}\times \bcube{d_1}\to \cube{\ell}$ be a $(k,\epsilon)$ linear seeded extractor with seed length $O(\log(n/\epsilon))$ (this is for the challenge/final output, we use \Cref{thm:cglext})
   \STATE Let $\lext:\bcube{n}\times \bcube{d_2}\to \cube{\ell}$ be a $(k,1/n^2)$ linear seeded extractor (this is for the response, we use \Cref{thm:li_lext})
   \STATE Let $\mathrm{AffineExt}:\bcube{n}\to \{0,1\}$ be an affine extractor (\Cref{thm:affine_extractor}) 
   \STATE \hrulefill
    \begin{enumerate}
        \item Let $s=x_{[d_1]}$.
        \item Let $a=\cglext(x,s)$.
        \item For each $i\in \bcube{d_2}$, check if $a=\lext(x,i)$.
        \item If there exists an $i\in \bcube{d_2}$ that satisfied the previous check, then output $\mathrm{AffineExt}(x)$.
        \item Else, output $a_1$.
    \end{enumerate}
\end{algorithmic}    
\end{algorithm}
For each variable ($x,s,a$) defined in \Cref{alg:quadratic_disperser}, we use capitalized letters $X,S(X),A(X)$ etc to denote the corresponding random variables. That is, $S(X)=X_{[d_1]}$, $A=\cglext(X,S(X))$ and so on for random variable $X$. 

We analyze the disperser by splitting into two cases based on if $X$ is $n^{0.2}$-pure.
\subsection{$X$ is not Pure}
\begin{theorem}
\label{thm:qdisp_affine}
    Let $X=f(U_s)$ be a quadratic source with an affine subsource $W$ with min-entropy at least $n^{0.2}$. Then $\qdisp$ (from \Cref{alg:quadratic_disperser}) is a disperser for $X$.
\end{theorem}
\begin{proof}
    We will restrict our attention to $W$ and show that $\qdisp$ is a disperser for $W$. A disperser on a subsource is also a disperser on the original source. 
    
    Since $S(W)$ has output length at most $d_1=100n^{0.1}$, fixing $S(W)=s$ only reduces the dimension by $d_1$. Thus, $W$ has a subsource $W'$ of dimension at least $\Omega(n^{0.2})$ where $S(W')=s$ is a constant. Now, note that $A(W')$ is a linear function of $W'$ (as $s$ is fixed). Thus, we can further restrict $A(W')=a$ at a loss of additional dimension $\ell=0.1n^{0.1}$ to obtain a subsource $W''$ of dimension $\Omega(n^{0.2})$ where $A(W'')=a$ is a constant. 
    
    From \Cref{lem:rao_affine_lext}, we have that there exists $i\in \bcube{d_2}$ such that $\lext(W'',i)=U_\ell$. Condition on the event that $\lext(W'',i)=a$. This happens with probability at least $2^{-\ell}$ over $W''$. Furthermore, this conditioning reduces the dimension of $W''$ by at most $\ell\leq 0.1 n^{0.1}$. Let $\tilde{W}$ be the subsource of $W''$ obtained after this conditioning. This subsource has min-entropy at least $\Omega(n^{0.2})$. Restricted to the subsource $\tilde{W}$, the condition checked in Line~3 of \Cref{alg:quadratic_disperser} holds with probability $1$ for the seed $i$ identified earlier. Thus, on $\tilde{W}$, the output distribution of the disperser is $\affinext(\tilde{W})$ which is $0.1$ close to $U_1$ and hence is not constant. This completes the proof. 
\end{proof}

\subsection{$X$ is Pure}

We are now ready to argue that $\qdisp$ is an extractor for pure quadratic sources.
\begin{theorem}
\label{thm:qdisp_nonaffine}
Let $X$ be a $n^{0.2}$-pure quadratic source with $\hinf(X)\geq n-n^{0.1}$. Then, $\qdisp$ is an extractor for $X$ with error $2^{-\Omega(n^{0.1})}$. 
\end{theorem}
\begin{proof}
Let $\Delta=n^{0.1}$ and let $\hinf(X)\geq n-\Delta$.
    From \Cref{lem:input_reduction_entropy}, we have that $X$ is $\epsilon=2^{-\Omega(n^{0.1})}$ close to a convex combination of quadratic subsources $X'$ where $\hinf^{\epsilon}(X')\geq n-\Delta-1$ and the generating map $f$ of $X'$ has seedlength $m=n-\Delta+O(n^{0.1})$.   It suffices to prove that $\qdisp$ extracts from such subsources $X'$. By \Cref{lem:purity_preserving}, these sources are still $n^{0.2}$-pure. Henceforth, we will use $X$ to refer to $X'$. Thusm $X$ is a quadratic source with generating map $f$,  $m=n-\Delta+O(n^{0.1})$ generating bits and satisfying $\hinf^{\epsilon}(X)\geq n-\Delta-1$. 

    We use $V=U_m$ to denote the input seed for the source. That is, $X=f(V)$.
    From our choice of $d_1=100\Delta$ and \Cref{lemma:smooth_entropy_slice}, we have that $\hinf^{\epsilon}(S(X))\geq 0.99 d_1$. 

    \paragraph{Quadratic conditioning: input reduction}
Some additional set up will be useful. Since restricting to a subset of bits is a linear operation, we have that $S=h(V)$ for a quadratic map $h$.  From \Cref{lem:input_reduction_short}, we have a full rank linear map $L:\bcube{m}\to \bcube{m-r}$ such that 
\begin{equation}
\label{eqn:dist_tv_input_reduction}
\dtv((h(V),L(V)),(h(V),U'_{m-r}))\leq \epsilon
\end{equation} where $U'$ is a uniform string independent of $V$ and $r=100\Delta+O(\log (1/\epsilon))$. Let $e_1,\ldots,e_{r}$ be a basis for the nullspace of $L$. Let $e_{r+1},\ldots,e_{m}$ be a completion of the basis to span all of $\bcube{m}$. We have that $V=A+B$  $A$ is uniform over the span of $e_{r+1},\ldots,e_m$, $B$ is uniform over the nullspace of $L$ and $A$ and $B$ are independent. Observe that we can express $A=M_1(U_{m-r})$ and $B=M_2(U_{r})$ for appropriate full rank linear maps $M_1$ and $M_2$. We will use $Y=U_{m-r}$ and $Z=U_{r}$ to denote the input random variables of $M_1$ and $M_2$. Thus, the original source can be rewritten as $X=f(M_1(Y)+M_2(Z))$. Also, $L(M_1(Y)+M_2(Z))=L(M_1(Y))$. From our choice of $M_1,M_2,Y,Z$, we have that $L(M_1(Y))$ uniquely determines $Y$. Thus, \Cref{eqn:dist_tv_input_reduction} now becomes
\begin{equation}
    \label{eqn:dist_tv_input_reduction_final}
    \dtv(((h(Y,Z)),Y),(h(Y,Z),U_{m-r}))\leq \epsilon
\end{equation}
where $U_{m-r}$ is independent of $(Y,Z)$. Henceforth, we will rename the map $(Y,Z)\mapsto f(M_1(Y)+M_2(Z))$ as $f(Y,Z)$. The above argument has established the following claim.
\begin{claim}
\label{claim:s_indep_z}
    For $Y,Z$ defined as above, we have that
    \[
    \dtv((S(X)\mid Y=y),S(X))\leq O(\sqrt{\epsilon})
    \]  
    with probability $1-\sqrt{\epsilon}$ over $y\sim Y$.
\end{claim}
We first show the following lemma.
    \begin{lemma}
        \label{lem:A is uniform}
        The distribution of $A(X)$ is $2^{-\Omega(n^{0.1})}$-close to $U_\ell$.
    \end{lemma}
    \begin{proof}
        Recall that $S(X)=h(Y,Z)$ where $h$ is a quadratic map with output length $d_1=100\Delta$. Expand $h$ as $h(Y,Z)=h_1(Y)+h_2(Y,Z)+h_3(Z)$ where $h_2$ is a bilinear map. Let $\ell_1,\ldots \ell_j$ be the linear functions of $Y$ that are used in $h_2$. Clearly $j\leq 100\Delta\cdot r\leq O(n^{0.2})$. 
        \paragraph{Quadratic conditioning: Cohen-Tal conditioning}
        Our goal is to fix functions of $Y$ so that $S$ is only a function of $Z$. We want to do this fixing without affecting the distribution of $S$, and also retaining entropy in $Y$. \Cref{claim:s_indep_z} allows us to achieve the former and the fact that $S$ is a short quadratic map allows us to achieve the latter.
        
        First, we consider the subsource of $X$ obtained by fixing the value $\ell_1,\ldots \ell_j$. With probability $1-\epsilon^{1/4}$ over this fixing, the subsource $X'=f(Y',Z)$ has $S(X')=h(Y',Z)=h_1(Y')+h_3(Z)$ ($h$ has no bilinear terms that depend on $Y'$) and $Y'=M(U_{m-r-j})$ for some full rank linear map $M$. Absorbing the linear map $M$ into the source and overloading notation (redefining $f,h,Y\gets M^{-1}(Y')$), we have that the subsource we reached is of the form $X=f(Y,Z)$ with $S(X)=h(Y,Z)=h_1(Y)+h_3(Z)$ and $\hinf^{2\epsilon^{3/4}}(X)\geq n-\Delta-j-O(\log(1/\epsilon)$. Furthermore, $Y=U_{m-r-j}$.

        From \Cref{clry:cohen_tal}, we have a function $\ctpartition:\bcube{m-r-j}\to \bcube{m-r-j-\Omega((m-r-j)/d_1)}$ such that $h_1$ is constant on $\ctpartition^{-1}(b)$ for any $b$.
        With  probability $1-O(\epsilon^{1/4})$ over fixing $\ctpartition(Y)$, we land in a subsource of the form $X'=f(Y',Z)$ with $\hinf^{O(\sqrt{\epsilon})}(X')\geq c(m-r-j)/{d_1})+r-O(\log(1/\epsilon))$ for some constant $c>0$. On this subsource, we have $S(X)=h_3(Z)$. Let $\overline{k}=\Omega((m-r-j)/d_1)$. From our choice of parameters, we have that $\overline{k}= \Omega(n^{0.9})$. Redefining the notation, our final source has the structure $X=f(Y,Z)$ with (1) $S(X)=h_3(Z)$, (2) $Y=U_{\overline{k}}$, (3) $\hinf^{O(\sqrt{\epsilon})}(X)\geq \overline{k}+r-O(\log(1/\epsilon))$. 

        Also, since we only fixed functions of $Y$ throughout the conditioning, \Cref{claim:s_indep_z} implies that $\hinf^{O(\sqrt{\epsilon})}(S(X))\geq 0.99d_1$. 
        \paragraph{Structure theorem}
        Now, we apply \Cref{clry:pure_quadratic_structure} to $X$. Recall that  $X$ is $n^{0.2}$-pure. We set $s=\overline{k},t=10n^{0.2},r=n^{0.6}$ and entropy parameter $k=\overline{k}+|Z|-O(\log(1/\epsilon))$. Thus, we have that $X$ is $2^{-\Omega(n^{0.1})}$-close to a convex combination of subsources of the form 
        \[
        X'=g(Y',Z)=g_1(Y')+g_2(Y',Z)+g_3(Z)
        \]
        such that (1) $\min_{a\neq 0}\hinf(\partial_a g_1(.))\geq n^{0.6}$ and (2) $Y'=U_{s'}$ for $s'\geq \overline{k}/2$. Importantly, $Z$ is left unchanged.

        Since $S(X)$ depends only on $Z$, its distribution is left unchanged. We can now apply \Cref{thm:cauchy_schwarz_general} with $k=\Omega(n^{0.6})$, $\epsilon=2^{-\Omega(n^{0.1})}$, $d=d_1=O(n^{0.1})$ and $\Delta=n^{0.1}$. We use the linear seeded extractor from \Cref{thm:cglext} (we truncate the output to $0.1n^{0.1}$ bits). Note that the entropy requirement of $\cglext$ is satisfied as $r\geq n^{0.6}\gg \log(n/\epsilon)$. Thus, we have that $A(X)$ is $2^{-\Omega(n^{0.1})}$-close to $U_\ell$ for all the subsources in the convex combination. Adding up all the TV distances due to various conditionings, we still have that that $A$ is $2^{-\Omega(n^{0.1})}$ close to $U_\ell$. This completes the proof of the lemma.
     \end{proof}

    \begin{lemma}
    \label{lem:low_collision_probability}
        The probability that there exists an $i\in \bcube{d_2}$ such that $A(X)=\lext(X,i)$ is at most $2^{-\Omega(n^{0.1})}$.
    \end{lemma}
    \begin{proof}
    Throughout the proof, we will set $\epsilon=2^{-\Omega(n^{0.1})}$
    The proof uses the same idea as the proof of \Cref{lem:A is uniform}. We will argue that $\Pr[A(X)=\lext(X,i)]\leq 2^{-\Omega(n^{0.1})}$ for a fixed $i$. The final claim follows by a union bound on $\poly(n)$ choices of $i$.  

    Again, we assume that $X=f(Y,Z)$ with $S(X)=h(Y,Z)$ satisfying \Cref{claim:s_indep_z}. Since $\lext$ is a linear seeded extractor, note that $\lext(X,i)$ is also given by a quadratic map $g(Y,Z)$ which has $\ell=0.1n^{0.1}$ output bits. We first write $g(Y,Z)=g_1(Y)+g_2(Y,Z)+g_3(Z)$. Our aim is to fix functions of $Y$ to make both $S(X)=h(Y,Z)$ and $\lext(X,i)=g(Y,Z)$ solely functions of $Z$. 
    
    Note that one can treat $(h,g)$ as a quadratic map with $d_1+\ell\leq O(n^{0.1})$ output bits and apply the quadratic conditioning argument again. After conditioning on appropriate functions of $Y$, the source obtains the form $X=f(Y,Z)$ with $Y=U_{\overline{k}}$, $\overline{k}\geq n^{0.9}$, $S(X)=h_3(Z)$ and $\lext(X,i)=g_3(Z)$. We now again repeat the application of the structure theorem (\Cref{clry:pure_quadratic_structure}) as in the proof of \Cref{lem:A is uniform} to prepare the source to be of the form required by \Cref{thm:cauchy_schwarz_general}. The proof is then completed by applying \Cref{thm:cauchy_schwarz_general}. 
    
    \Cref{thm:cauchy_schwarz_general} guarantees that with high probability over fixing $z\sim Z$, the function $A=\cglext(f(Y,z),h_3(z))$ looks $2^{-\Omega(n^{0.1})}$ close to $U_\ell$. However, note that fixing $Z$ fixes $\lext(X,i)$ (as we had earlier made $\lext(X,i)$ solely a function of $Z$). Thus, after fixing $Z$, we have that $A$ is close to uniform and $\lext(X,i)$ is constant. This implies that the probability that $A=\lext(X,i)$ is at most $2^{-\ell}+2^{-\Omega(n^{0.1})}\leq 2^{-\Omega(n^{0.1})}$. Now, applying a union bound over $i\in \bcube{d_2}$ completes the proof.
    \end{proof}
    Given, the two lemmas above, the proof that $\qdisp$ is an extractor is straightforward. First, \Cref{lem:low_collision_probability} implies that with probability $1-2^{-\Omega(n^{0.1})}$, execute Step~5 of \Cref{alg:quadratic_disperser} and output $A_1$. \Cref{lem:A is uniform} establishes that this bit is $2^{-\Omega(n^{0.1})}$ close to  uniform. 
\end{proof}
We now complete the proof of the main theorem \Cref{thm:main_disp}.
\begin{proof}[Proof of \Cref{thm:main_disp}]
    The theorem follows immediately by combining \Cref{thm:qdisp_affine} and \Cref{thm:qdisp_nonaffine}.
\end{proof}
\section{Affine Extraction}
\label{sec:qext_is_affine_extractor}
We first show that the construction in \Cref{alg:quadratic_correlation_breaker} is also an affine correlation breaker (see \Cref{defn:affine_correlation_breaker}). This argument is not new to our paper, it was already sketched in Section~2.3 of \cite{chattopadhyay2022extractors} and we include a formal proof for completeness.
\begin{theorem}
\label{thm:affine_correlation_breaking}
    For any $\delta\in (0,1)$ and positive integers $t,a,m$, the function $\qcb$ (from \Cref{alg:quadratic_correlation_breaker}) with parameters instantiated the same way as in \Cref{thm:quad_correlation_breaking} is also a $(t,k,O(\delta))$-affine correlation breaker.
\end{theorem}
\begin{proof}
    Our proof is much easier than the quadratic analogue in \Cref{thm:quad_correlation_breaking}. We first do a step of affine conditioning. 
    \paragraph{Affine conditioning} Recall that $W,W^{1},\ldots,W^{t}$ all have exactly $t$ output bits and are linear functions of the seed bits $V$. From \Cref{lem:affine_conditioning} (applied to $\ell,\ell^{[t]}$, we have that there exists subspaces $A,B$ such that (1) $A+B=V=\F_2^{k}$, (2) $\dim(B)\leq d(t+1)$, (3) $\dim(A)\geq k-d(t+1)$ and (4) $\ell(A),\ell^1(A),\ldots,\ell^{t}(A)$ is constant over the support of $A$.

    Recall initially we have that $X=f(V)=c+M(V)=c+M(A)+M(B)$ for constant vector $c$ and linear map $M$. After the conditioning above, we have that $S_1,S_1^{[t]}, W,W^{[t]}$ are solely functions of $B$. Thus, we have that 
    \begin{equation}
    \label{eqn:affine_CB_R}
R=\lext(c+M(A)+M(B),S_1(B))=\lext(c+M(A),S_1(B))+\lext(M(B),S_1(B)).
    \end{equation}
    Since $M$ is a full rank map and $\dim(A)\geq k-d(t+1)$, we have that $c+M(A)$ satisfies the entropy requirement of $\lext$. Also, $S_1(B)$ is uniform. Thus, we have that with probability $1-\delta$ over fixing $S_1(B)$, it holds that $R$ is uniform. In fact, we can also fix $S_1^{[t]}(B)$ without affecting the uniformity of $R$. After this fixing we have that $\hinf(W)\geq d-(t+1)d_0-O(\log(1/\epsilon))$ (with probability $1-\delta$ over the fixing) from min-entropy chain rule.

    Now, we make $R,R^{[t]}$, independent of $W,W^{[t]}$. To do that we have to also fix the second term in \Cref{eqn:affine_CB_R} and the analogous terms in $R^{[t]}$. These are solely functions of $B$, so fixing them retains the uniformity of $R$. Since $\lext$ outputs $d_1$ bits, fixing these functions of $B$ makes the entropy of $W$ now  $d-(t+1)(d_0+d_1)-O(\log(1/\delta)$ with probability $1-O(\delta)$ over all the fixings. At this point, we have that $\{R,R^{[t]}\}$ is independent of $\{W,W^{[t]}\}$, $R$ is uniform and $W$ satisfies the entropy requirement of $\CB$. Now, we apply to correlation breaking guarantee \Cref{thm:li_correlation_breaker} to complete the proof.
\end{proof}

We are now ready to complete the proof of \Cref{thm:qext_is_affine_extractor}.

\begin{proof}[Proof of \Cref{thm:qext_is_affine_extractor}]
    Let our affine source be $X=f(V)$ for $V=U_{k}$,
We will first argue that the random variable $W=(\rcombext(X,i))_{i\in [N]}$ has the following property: $W_i= U_d$ with probability $1-1/n^2$ over $i\sim [N]$. This follows from Item~2 in \Cref{clry:explicit_rank_combiner_extractor} and \Cref{lem:rao_affine_lext}.  We also have that $W_i=\ell^i(V)$ for some affine functions $\ell^{i}$.

Let $Q\subseteq [N]$ be the set of indices where $W_{i}=U_d$. Consider the random variable \[S=(\qcb(X,W^i,i))_{i\in [N]}.\] We will argue that for all sets $T\subseteq Q$ with $|T|\leq t$, we have that $|\E[\chi_T(S)]|\leq O(\delta)$.
This is immediate from \Cref{thm:affine_correlation_breaking} and the fact that $X,W$ satisfy the premise (by our choice of parameters) and the indices $i\in [N]$ form valid advice strings. 

Now, applying \Cref{fact:agm} implies that $S$ is $N^{-t}$-close to a $(q,t)$-NOBF source with $q=N/n^2$ and $t=(m\log n)^{C_2}$ chosen earlier. The theorem statement now follows from \Cref{thm:nobf_extractor}.
\end{proof}
\section{Additional Proofs}
\label{sec:additional_proof}
We give the proof of the probabilistic construction of the rank combiner.
\begin{proof}[Proof of \Cref{thm:rank_combiner_probabilistic}]
Let $G_1,\ldots,G_{n}\in \F_2^{k\times k}$ such that $\dim\{G_i a\}_{i\in [n]}\geq d$ for all $a\neq 0$. Let $r$ be the target rank, we fix it later. We first argue that a uniform random matrix $\mathcal{W}\in \F_2^{m\times n}$ satisfies 
\begin{equation}
\label{eqn:rank_combiner_guarantee_probabilist}
\rank_{\F_2}\left(\sum_{i=1}^{n}(c^{\top}\mathcal{W})_iG_i\right)\geq r
\end{equation}
for all non-zero $c\in \F_2^{m}$ with high probability over $\mathcal{W}$ . Indeed for each $c\neq 0$, the vector $c^{\top}W$ is a random vector in $\F_2^{n}$. Thus, it suffices to argue that a random vector $X$ satisfies 
\begin{equation*}
\rank_{\F_2}\left(\sum_{i=1}^{n}X_iG_i\right)\geq r
\end{equation*}
with high probability over $X$ and then union bound over non-zero $c$.

This is what we will do. Lower bounding the rank of a matrix is equivalent to upper bounding the size of its null space. For any $a\neq 0$, we have that the matrix $G_a=\{G_i a\}_{i\in [n]}$ has rank at least $d$. That means with probability at most $2^{-d}$, it holds that $G_aX=0$ for uniform random vector $X$. Thus, we have that 
\[
\E_X[|\{a\mid (\sum_{i=1}^{n}X_iG_i) a=0\}|]\leq \sum_{a\in \F_2^{k}}\Pr[G_aX=0]\leq (2^{k}-1)/2^{d}.
\]
By Markov's inequality, with probability at least $1-\delta$, we have $\rank_{\F_2}\left(\sum_{i=1}^{n}X_iG_i\right)\geq d-\log(1/\delta)$. Union bounding over $c$, we have that with probability at least $1-\delta\cdot 2^{m}$, \Cref{eqn:rank_combiner_guarantee_probabilist} holds. Choosing $\delta=\epsilon/2^{m+1}$ and $r=d-(m+1)-\log(1/\epsilon)$, we have that \Cref{eqn:rank_combiner_guarantee_probabilist} holds with probability $1-\epsilon/2$. Now sample $N$ independent random matrices $W_1,\ldots, W_N$ for $W=O(nk^2/\epsilon)$. 
Note that the total number of choices of $G_1,\ldots, G_n$ is atmost $2^{nk^2}$. Thus, by a Chernoff bound and the bound we just established, we have that $W_1, W_2,\ldots, W_N$ is a $(d,d-m-\log(1/\epsilon),\epsilon)$-rank combiner with $N=O(nk^2/\epsilon)$.
\end{proof}

\end{document}